\documentclass[manuscript]{acmart} 
\pdfoutput=1

\AtBeginDocument{%
  \providecommand\BibTeX{{%
    \normalfont B\kern-0.5em{\scshape i\kern-0.25em b}\kern-0.8em\TeX}}}

\setcopyright{acmcopyright}
\copyrightyear{2022}
\acmYear{2023}
\acmDOI{10.1145/1122445.1122456}

\acmConference[]{}{}{}
\acmBooktitle{}
\acmPrice{15.00}
\acmISBN{}

\usepackage{subfigure} 
\newcommand{\red}[1]{\textcolor{red}{#1}}

\begin{document}

\title{Fields: Toward Socially Intelligent Spatial Computing}


\author{Leonardo Giusti}
\email{lgiusti@google.com}
\affiliation{%
  \institution{Google}
  \streetaddress{1600 Amphitheatre Parkway}
  \city{Mountain View}
  \state{California}
  \postcode{94043}
}

\author{Lauren Bedal}
\email{lbedal@google.com}
\affiliation{%
  \institution{Google}
  \streetaddress{1600 Amphitheatre Parkway}
  \city{Mountain View}
  \state{California}
  \postcode{94043}
}

\author{Eiji Hayashi}
\email{eijihayashi@google.com}
\affiliation{%
  \institution{Google}
  \streetaddress{1600 Amphitheatre Parkway}
  \city{Mountain View}
  \state{California}
  \postcode{94043}
}

\author{Timi Oyedeji}
\email{oyedeji@google.com}
\affiliation{%
  \institution{Google}
  \streetaddress{1600 Amphitheatre Parkway}
  \city{Mountain View}
  \state{California}
  \postcode{94043}
}

\author{Jin Yamanaka}
\email{yamanakaj@google.com}
\affiliation{%
  \institution{Google}
  \streetaddress{1600 Amphitheatre Parkway}
  \city{Mountain View}
  \state{California}
  \postcode{94043}
}

\author{Colin Bay}
\email{colin.bay@concreteux.com}
\affiliation{%
  \institution{Concrete UX}
  \streetaddress{2189 NW Wilson St}
  \city{Portland}
  \state{Oregon}
  \postcode{97210}
} 

\author{Ivan Poupyrev}
\email{ipoupyrev@google.com}
\affiliation{%
  \institution{Google}
  \streetaddress{1600 Amphitheatre Parkway}
  \city{Mountain View}
  \state{California}
  \postcode{94043}
}


\begin{abstract}
In our everyday life, we intuitively use space to regulate our social interactions. When we want to talk to someone, we approach them; if someone joins the conversation, we adjust our bodies to make space for them. In contrast, devices are not as considerate: they interrupt us, require us to input commands, and compete for our attention. In this paper, we introduce Fields, a design framework for ubiquitous computing that informs the design of connected products with social grace. Inspired by interactionist theories on social interaction, Fields builds on the idea that the physical space we share with computers can be an interface to mediate interactions. It defines a generalized approach to spatial interactions, and a set of interaction patterns that can be adapted to different ubiquitous computing systems. We investigated its value by implementing it in a set of prototypes and evaluating it in a lab setting.

\end{abstract}

\ccsdesc[500]{Human-centered computing~Implicit input}

\keywords{ubiquitous computing; ambient computing; radar sensing; implicit interaction; proxemics}

\begin{figure}
\centering
  \includegraphics[width=1\columnwidth]{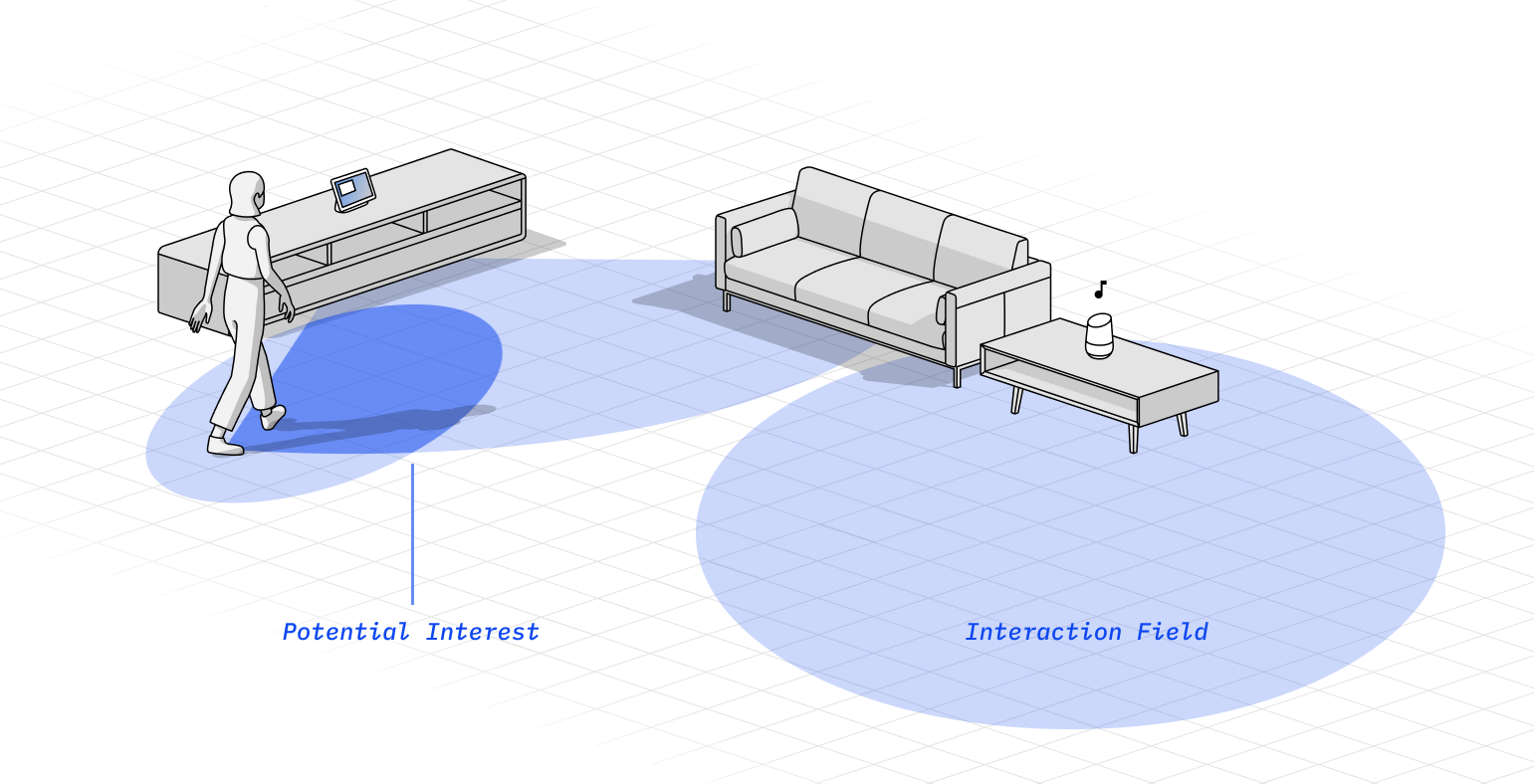}
  \caption{We introduce Fields, a design framework for ubiquitous computing systems. This approach defines interaction fields as regions of space around users and devices that capture a metric of Potential Interest (PI). This measures the amount of engagement between a user and device based on the overlap between interaction fields. 
 \red{}}~\label{fig:rtr_hero}
\end{figure}

\maketitle
\section{Introduction}
Ubiquitous computing \cite{weiser1991computer, abowd2000charting, gershenfeld1999things} is here. Thermostats, doorbells, and speakers all have sensing capabilities, process information, and run machine learning algorithms. The average U.S. household has 25 connected devices, more than double the amount from 2019 \cite{koetsier_2022}. Computers have proliferated our spaces —however, they are far from being invisible. 

Pioneer Mark Weiser’s vision for ubiquitous computing described a future where devices support us while we are engaged in the physical world without spending our attention \cite{weiser1997coming}. Instead, connected products seem to do the opposite. First, they often interrupt our everyday activities \cite {anderson2018survey, gibbs2005considerate}, diverting our attention with notifications and alerts. The effect of this information being communicated irrespective of a user's current context is negative, in terms of task completion \cite {bailey2006need}, performance \cite {leroy2009so}, and emotional state \cite {mark2008cost}. Furthermore, devices today frequently demand input \cite{gershenfeld1999things, ju2008design} and rely on explicit input methods, predominately voice or touch. For example, products today require using a wake word —e.g. “OK Google” to begin a conversation with a voice assistant; a double-tap is often required to activate a smart display. The repetition of these commands is tiresome \cite{ahire2020tired}, and hinder the natural flow of a conversation \cite{clark2019makes}. These demands of user's input and attention prevent the seamless integration of products into our everyday life in a way that feels considerate and effortless. 

At the core of these concerns lies two key challenges for ubiquitous computing: context-awareness  \cite{lyytinen2004surfing, hong2009kim, dey2001understanding, lyytinen2002ubiquitous}, and interaction techniques \cite{abowd2002human,ju2008design, greenberg2011proxemic}. 

Context-awareness is often approached as a technical challenge requiring applications, as Gregory Abowd and Elizabeth Mynatt describe, "to adapt their behavior based on information sensed from the physical and computational environment. Many applications have leveraged simple context, primarily location and identity, but numerous challenges remain in creating reusable representations of context, and in creating more complex context from sensor fusion and activity recognition." \cite{abowd2000charting}. The research on interaction technique represents another view: it suggests that interactions can be designed to be far less jarring or disruptive than what we experience now. In the context of ubiquitous computing, many complimentary schools of thought have emerged regarding the need of interaction techniques that facilitate interactions including natural interfaces \cite{abowd2000charting}, considerate computing \cite{gibbs2005considerate}, attentive user interfaces \cite{vertegaal2003attentive}, and calm technology \cite{weiser1996designing}. 

This paper focuses on the latter, suggesting new paradigms for interaction as computers move "off-desktop" and into the physical world. We focus our investigation on the idea of natural interfaces, defined here as  interaction methods that understand our implicit, nonverbal cues in order to unlock a rich method of communication between people and devices. As Wendy Ju eloquently describes, "'Implicit interactions' - those that occur without the explicit behest or awareness of the user- will become increasingly important as human-computer interactions extend beyond the desktop computer into new arenas" \cite{ju2008design}. 

Implicit communication is natural to us as humans. We understand each other intuitively, often without an explicit command \cite{clark1996arranging, goffman1967interaction}: we hold the door open for others when we see that their arms are full, we keep certain distances to others depending on familiarity; when we want to talk to someone, we approach them; if someone else joins the conversation, we adjust our bodies to make space for them. Nonverbal cues such as changes of spatial relationships are an implicit form of communication that we understand intuitively, and allows us to seamlessly initiate, maintain or end a conversation with others –what if computers could understand us this way?

In this paper, we introduce "Fields" a framework to design ubiquitous computing systems with social grace, which supports the design of implicit interactions that feel considerate, welcoming, and reciprocal \cite{ju2008design}. Inspired by social behaviors and interactionist theories by Adam Kendon \cite{kendon1990conducting} and Erving Goffman \cite {goffman1967interaction} this approach focuses on one core idea: the physical space we share with computers can be used as an interface to mediate interactions. 

\begin{itemize}
\item First, we introduce two new concepts, \textit{Interaction Fields} and \textit{Potential Interest}, as key components of the Field framework. An \textit{Interaction Field} is a region of space surrounding a user or a device that carries a particular significance for interactions; the \textit{Potential Interest} tracks the overlap between Interaction Field to continuously estimate the level of engagement between people and devices. These concepts are defined independently of a specific technological implementation, can be applied to a variety of connected products and devices. 
\item Second, we demonstrate how the Potential Interest can be used to define a set of interaction patterns that solve typical interaction problems for ubiquitous computing systems, such as initiate and terminate interactions, turn-taking, etc.  These interaction patterns include Greeting, Turn-taking, and Revealing.
\item Third, we implemented four prototypes using these interaction patterns to explore how the Fields approach could be deploy in real-world scenarios.
\item Fourth, we conducted a study with 14 participants to evaluate a set of 4 use cases. Through self-reported measures on effort and focus as well as qualitative findings, our evaluation demonstrates receptiveness and desirability for using the proposed interaction technique as opposed to existing techniques.
\end{itemize}

\section{Related Work}
The majority of works to understand spatial relationships for ubiquitous computing originate from the idea of proxemics, a term coined by cultural anthropologist, Edward Hall \cite{hall1966hidden}. Proxemics addresses how people perceive, understand, and use space with an emphasis on varying distances between people. His book The Hidden Dimension, published in 1966, presented foundational ideas of how surrounding distances of people correlated with interpersonal relationships. Hall identified four types of informal space: (1) intimate, (2) casual—personal, (3) social—consultative, and (4) public. 

\begin{figure}[h]
\centering
  \includegraphics[width=0.5\columnwidth]{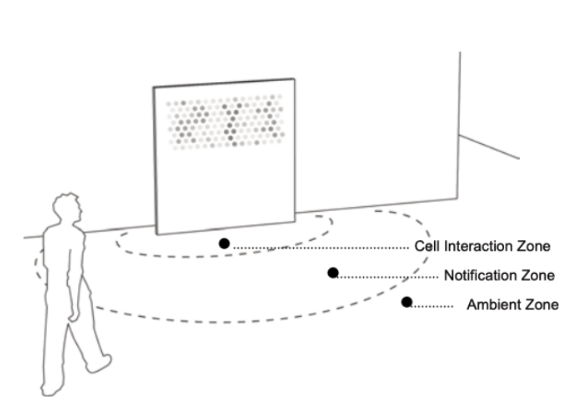}
  \caption{Hello.Wall \cite{prante2003hello}  leverages ideas of Hall’s proxemic theory by introducing the idea of ‘distance dependent semantics’, discrete zones around the display which allow for a progression from ambient information, to notifications, to direct interactions}
  \label{fig:hall}
\end{figure}

Hall's work is commonly seen within HCI. For example, Hello.Wall from 2003 (Figure \ref{fig:hall}) leverages ideas of Hall’s proxemic theory by introducing the idea of ‘distance dependent semantics’, discrete zones around the display which allow for a progression from ambient information, to notifications, to direct interactions \cite{ballendat2010proxemic}. Vogel et al.  \cite{vogel2004interactive} and Ju \cite{ju2008range} also show applications of proxemic frameworks, showing how notifications and personal and private information can surface based on a user’s proximity to the device. \cite{roussel2004proximity} The Audience Funnel Framework introduced by Michelis and Miiller \cite{muller2010requirements} maps interaction states based on discrete distances and correlating user motivations.

Although proximity is a valuable tool to understand spatial relationships, this approach provides only a course understanding of spatial relationships. The majority of proxemic interfaces works described above define distance through fixed, binary values \cite{ju2008range, prante2003hello, vogel2004interactive}. A binary understanding of proximity (in or out of a given threshold) between a user and device does not capture a multitude of natural movements that occur around a device such as orientation and trajectories which, when added together with proximity, contain valuable and relevant information for designers. For example, if I'm close to a device, but just passing by, close proximity to a display should not elicit a device response.

In 2010, The Proxemic Interaction Framework introduced by Greenberg et al. addresses this issue. It used Hall’s ideas as a foundation to operationalize proxemics for ubiquitous computing with one key difference: recognizing spatial relationships as a continuous, rather than discrete value \cite{marquardt2015proxemic}. The authors recognized that using only binary, proxemic values is limited, and instead introduced different proximal dimensions (position, orientation, movement and identity) to help facilitate transitions between people and large displays from awareness to direct interaction. Many works leverage this framework and resulting Proximity Toolkit \cite{marquardt2011proximity} for interactive media players \cite{ballendat2010proxemic} , smart devices \cite{radle2014huddlelamp}, public displays \cite{wang2012proxemic} and even device to device interactions \cite{marquardt2012cross}. For example, the Proxemic Presenter \cite{greenberg2011proxemic} by Miaosen Wang tracks distance, orientation and identification around a device specifically for presentations on large displays. This work leverages cues such as orientation and proximity towards the display to show presenter notes and quick shortcuts (e.g. ability to skip slides).

The majority of work on spatial interactions takes inspiration from Hall's work on proxemics \cite{hall1966hidden}, which references the space surrounding individuals as territories to defend from intruders; he describes the ways people arrange themselves in space to optimize their own comfort. This means that as you step too close to someone, they may step away and give up their territorial space. However, outside this conceptual view, when we look broadly at the way people use space in their everyday life, a more complex and rich picture emerges. Take for instance, how one might initiate a conversation. When we want to talk to someone, we approach them; in response, they might take a couple of steps toward us, or turn around if they have no intention to talk to us. If someone else joins the conversation, we all adjust our bodies to make space for them. In the interaction with others, we don't only respond to variations in distance, orientation and movement, we proactively use the space around us to establish different social relationships and negotiate different level of engagements. 

Several researchers within the interactionism perspective \cite{stryker2006symbolic, blumer1986symbolic} in sociology including Adam Kendon and Erving Goffman \cite{persson2018framing} have tried to identify the underlying principles that regulate spatial arrangements in social encounters. For example, Goffman noticed that when people enter into a “focused interaction” they place themselves in a spatial-orientation arrangement such that each is facing inward around a shared space \cite {goffman1967interaction}; they create a common domain or arena where glances, gestures and words can be exchanged. 

\begin{figure}
\centering
  \includegraphics[width=0.25\columnwidth]{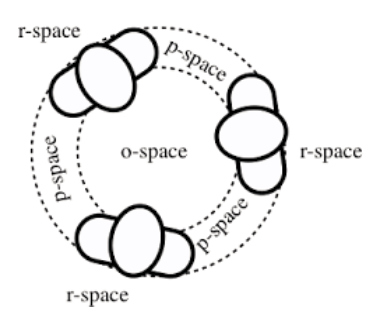}
  \caption{F-formations consist of 3 concentric spaces; the overlapping space between two people is called the O-space, a transactional segment that’s reserved for the main activity the group is pursuing. Other spatial areas which include the p-space, the ring of space surrounding a group of people, and the r-space which is the surrounding region, the outside world.}
  	\label{fig:o-space}
\end{figure}

In his book, Conducting Interaction \cite{kendon1990conducting}, Kendon describes the space for focused interaction as “transactional segments”, a space where individuals direct their attention and carry out activities. The size of this space can vary depending upon the activity in which they are engaged, e.g., watching television versus reading a book. It is also important to notice that a transactional segment is defined in relation to individuals’ lower body; thus by turning their head or shoulders, they can direct their gaze out of it. Kendon calls spatial arrangements, F-formations (or facing formation [18]). An F-formation is formed whenever two or more people arrange themselves such that their transactional segments overlap, creating a joint transactional space termed an o-space to which they have equal and exclusive access (Figure \ref{fig:o-space}). 

The interactionism perspective we believe provides a valuable foundation for the design of ubiquitous computing systems as it provides the conceptual tools to describe how people use space to initiate, maintain and terminate interactions with others. We can take inspiration from this sociological perspective to design connected products that are able to take initiative, and engage in interaction with users in a way that feels natural and considerate. 

\section{Fields Framework}
The goal of the Fields framework is to inform the design and development of ubiquitous computing systems that recognize and respect the user's intent to initiate, maintain and end interactions. 

The Field framework attempts to translate theories addressing the spatial relationships between people, particularly core concepts of transactional segements and social formations as described by Adam Kendon, into the domain of human computer interaction. We derived three principles to ground the creation of this framework, and the design of spatial interactions. 
\begin{itemize}
    \item \textbf{Coequal.} Both the user and the device play an equally important role in establishing the level of engagement within an interactive system, and their respective social spaces should be taken into consideration. 
    \item \textbf{Active.} The space surrounding an actor (either a device or a user) is not passively experienced as a defensive territory; rather, it is a social construct that can be used actively, to initiate, maintain, and end interactions. 
    \item \textbf{Emergent.} Interaction opportunities between an actor and a device emerge from the intersection of their individual spaces, and is continuously adjusted and negotiated. 
\end{itemize}

To facilitate the implementation of these principles in a variety of ubiquitous computing systems, the Fields framework defines two operative concepts: interaction fields, and potential interest.

\subsection{Interaction Field}
\begin{figure}
\centering
  \includegraphics[width=0.75\columnwidth]{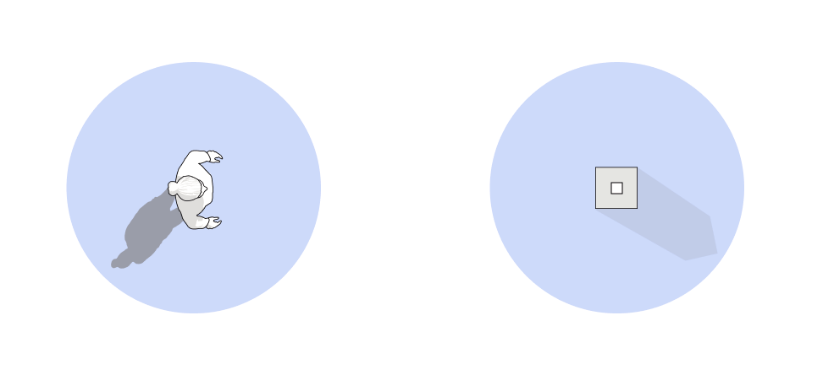}
  \caption{Interaction fields are regions of space around any ‘actor’ (person or device) within a computing system.}
  \label{fig:basic}
\end{figure}

In our framework, we define an \textit{interaction field} as the region of space surrounding an actor, either a user or a device, that carries particular significance for interaction (Figure \ref{fig:basic}).

The notion of “fields” has been used in other disciplines to define the peripersonal space \cite{teneggi2013social, bufacchi2018action}. The peripersonal space is the space surrounding the body where we can reach or be reached by external entities, including objects or other individuals \cite{rizzolatti1997space}. It represents a mediation zone between the body and the environment, that controls our interaction with the physical world as well as our social interactions. \cite{coello2021interrelation, lloyd2009space}. 

The shape of an interaction field can be specified by designers and developers to capture the unique characteristics of the involved actors, as well as the physical, cultural and social context where the interaction occurs. While Interaction Fields can assume a variety of shapes, we can identify two typical instantiations: (1) user's interaction field; (2) the device's interaction field.

\subsubsection{User interaction field}
\begin{figure}[h]
\centering
  \includegraphics[width=0.75\columnwidth]{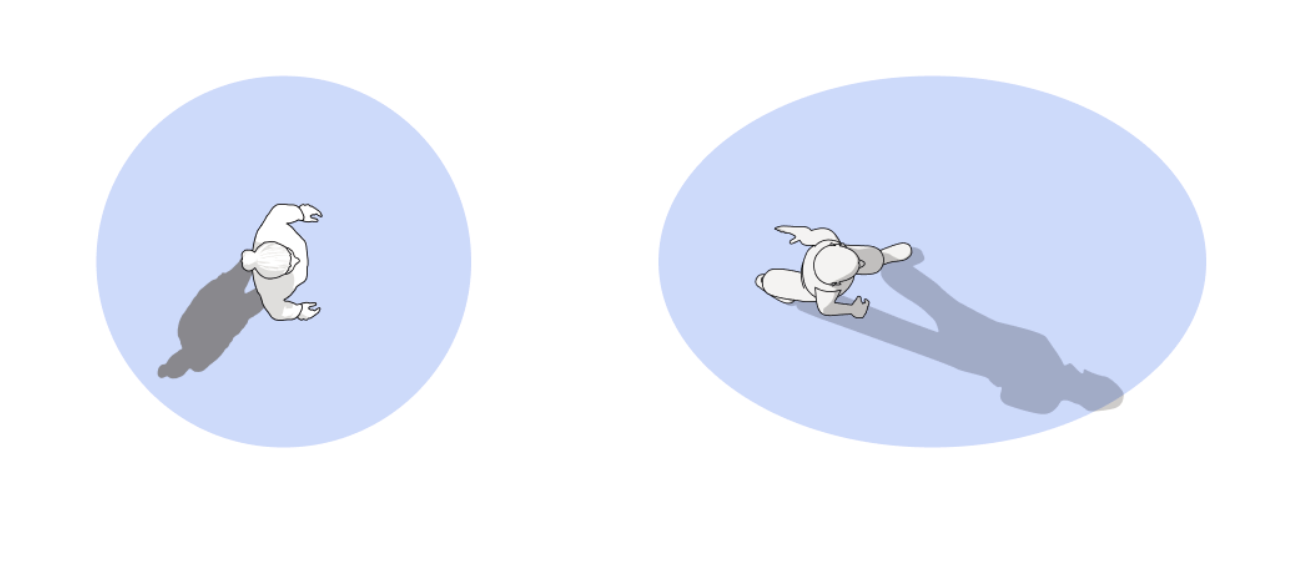}
  \caption{When the actor starts moving, the field morphs into an ellipsis with the eccentricity increasing as the speed of the actor increases. The overall area of the interaction field remains constant; as result, while the major axis of the ellipses increases as the speed increases, the minor axis of the ellipses decreases in length, projecting the interaction field forward.}
  \label{fig:user_interaction_filed}
\end{figure}

.  Fields conceptualizes an interaction field around a user, similar to the peripersonal space. In this paper, we started from using a circular shape for the ease of initial understanding and experimentation. Furthermore, as John Fruin mentions in his book, Designing for Pedestrians, the shape of personal space extends forward in the direction of the motion \cite{fruin1992designing}. This change of shape, especially when a user is in motion, contains valuable information about the actor’s intent and level of interest in their surrounding space. Based on these considerations, we define the the interaction field as an ellipsis, with the actor occupying one of the foci; the ellipses' major axis aligns with the actor direction of motion, and its eccentricity (distance between the foci) increases proportionally to the actor speed, while the overall area remains constant. When the actor is not moving, the shape of the interaction field is circular: since the actor velocity is equal to zero, the distance between the foci will be 0 as well. Based on existing literature around size of social spaces, we can approximate the radius of the interaction field (defining the size of the circle, when velocity is 0) to 1.2 meters \cite{hall1966hidden} (Figure \ref{fig:user_interaction_filed}).

\subsection{Device interaction field.} 
\begin{figure}
	\subfigure[Directional interaction field]{%
		\includegraphics[clip, width=0.40\columnwidth]{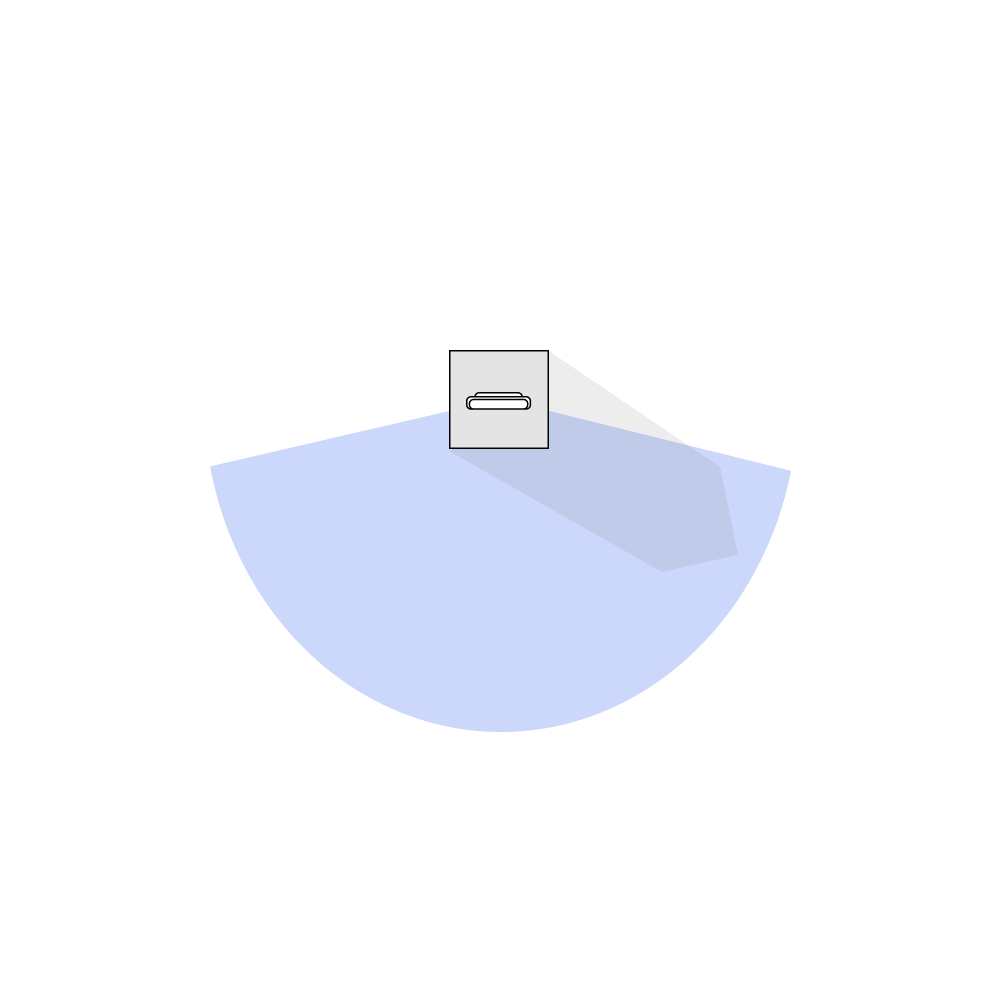}}%
	\hspace{2em}
	\subfigure[Non-directional interaction field]{%
		\includegraphics[clip, width=0.40\columnwidth]{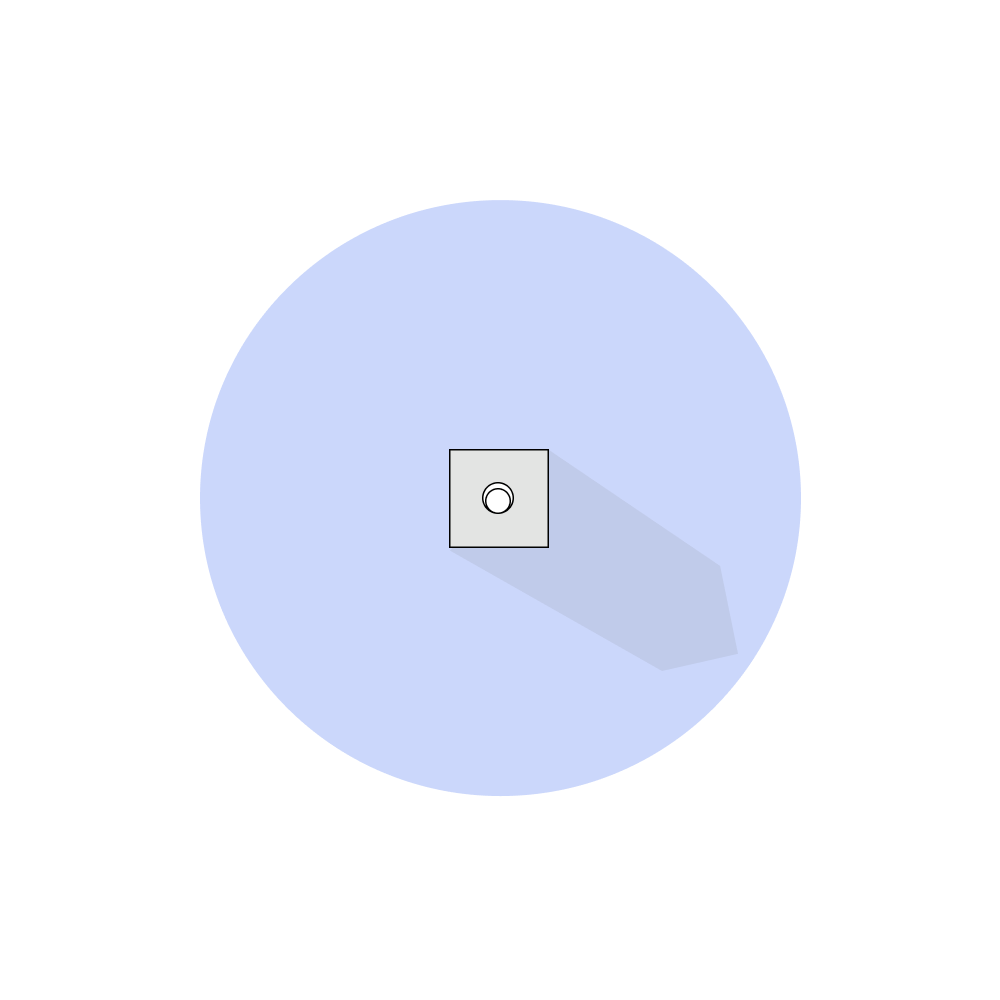}}%
	\caption{The non-directional orientation defines an interaction field that surrounds the device; while a directional orientation extends forward from one side of the device.} 
	\label{fig:device_field}
\end{figure}

The Fields framework also considers placing interaction fields around devices to give a sense of agency to both users and devices. Interaction fields around devices can vary in size; their configuration largely depends on technological constraints as well as the device functionality, context, input and form factors. For example, a TV will have very different interaction field than a smart speaker, since interactions with a TV will occur when a user is located in front of the device, while interactions with a smart speaker are independent of a user's angle relative to the speaker. Additionally, the device interaction field can support directional or non-directional sensing. The non-directional orientation defines an interaction field that surrounds the device; while a directional orientation extends forward from one side of the device. Non-directional orientation can be used when the angle of approach is not relevant for the interaction; directional orientation can be used to respond to users when they are in front of the device, rather than behind or to the sides. (Figure \ref{fig:device_field})

\subsection{Potential Interest}
The Potential Interest operationalizes the notion of social formations proposed by Kendon: as described in the previous section, when two or more people engage in a social interaction, they orient their transactional segments so that they overlap, creating a shared, communal space. Based on this consideration, we believe that the intersection between interaction fields contains valuable information about a user’s potential interest to engage in interaction. We define Potential Interest as a function of the overlap between two intersecting fields around a person and a device, which can be calculated as an IOU (intersection over union) value between the two fields (Eq \ref{eq:iou}) where $F_1$ and $F_2$ denote fields for the person and the device.

\begin{equation}
    \textnormal{Potential Interest} = \frac{F_1\cap F_2}{F_1\cup F_2}
    \label{eq:iou}
\end{equation}



\begin{figure}
	\subfigure[Potential Interest = 0.20]{%
		\includegraphics[clip, width=0.40\columnwidth]{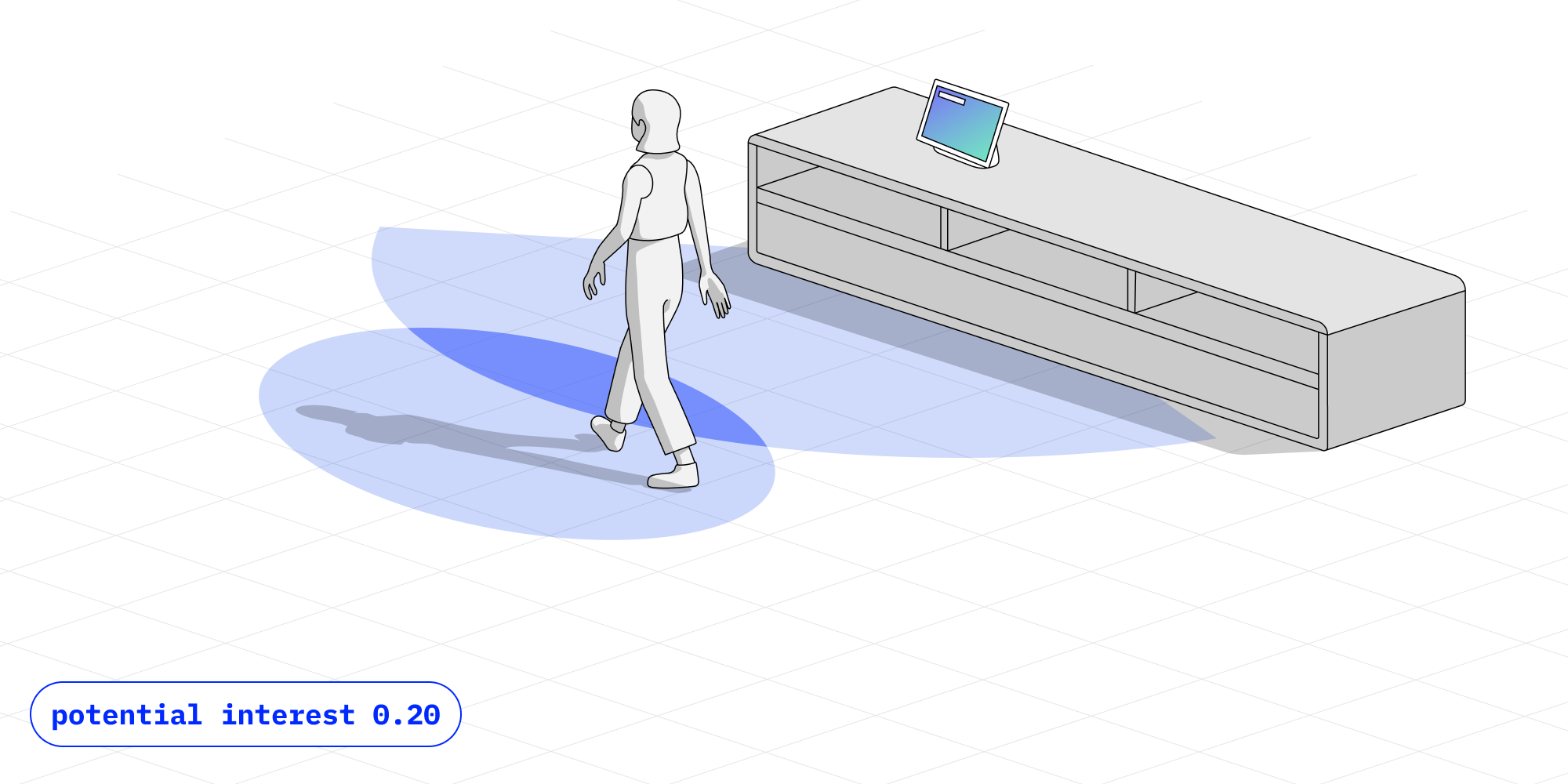}}%
	\hspace{2em}
	\subfigure[Potential Interest = .60]{%
		\includegraphics[clip, width=0.40\columnwidth]{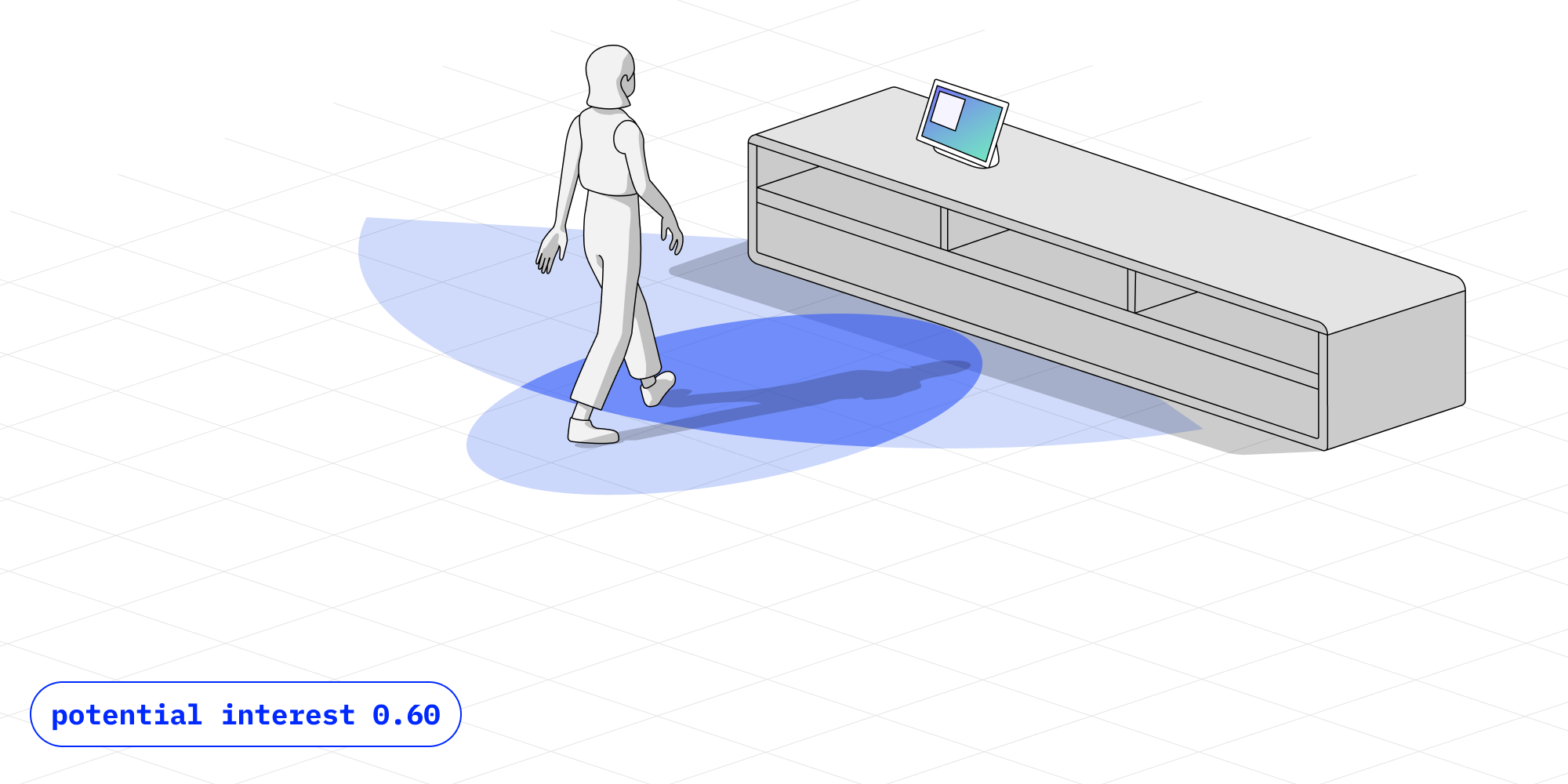}}%
	\caption{Additionally, even though a user is at a similar distance from the device, it could refrain from interrupting the user with a notification when a user is close but just walking by, and only reveal a notification once you’re engaged.} 
	\label{fig:pot_int_ex1}
\end{figure}

For example, when there is no overlap, the Potential Interest is equal to 0; when the user’s interaction field completely overlaps with the device’s interaction field, the Potential Interest is equal to 1. The potential interest allows to capture the complexities of the spatial relationship between people and devices, enabling ubiquitous computing system to respond to implicit interactions: for example, a user approaching a device will produce an increase of the potential interest that could trigger a change of state -e.g., the device shift from a sleep to an active state, revealing the content of a notification; however, if the user is just passing by, the device might refrain to interrupt the user. This simple example demonstrate how the potential interest captures important aspects of spatial relationships: in addition to distance, it takes into consideration orientation and movement (Figure \ref{fig:pot_int_ex1}).

.

\section{Interaction Patterns}

\subsection{Interaction Patterns}
Fields opens up the opportunity to define implicit interaction patterns, as reusable templates to address some of the common problems in the current generation of ubiquitous computing products (see Introduction). Each of these interaction patterns use the Potential Interest value as an input, and specify engagement thresholds to sequence interactions between users and devices. Thresholds are flexible parameters that specify the level of mutual focus (Potential Interest) required, for example, to trigger a certain device response, change device status (e.g. from sleep to active), or transition between implicit to explicit interactions. In this paper, we present three interaction patterns modeled around implicit behaviors we typically use in social interactions –Greeting, Turn-taking, and Revealing. 

\subsection{Greeting}
The Greeting pattern allows a user to wake-up or initiate a process on a device by intentionally overlapping its interaction field with the one of the device and engage in interaction without the use of explicit commands, such as hot-words or special gestures (e.g. double-tap on touchscreen). 
Previous work demonstrates that wake words and direct commands are frustrating to users \cite{ahire2020tired}. Moreover, as more devices are permeated around us, the disambiguation of a user’s focus to activate the right device is a critical problem \cite{fujinami2010interaction}. Particularly in our homes, where interactions with devices such as smart speakers and display are short and frequent, and often part of a larger set of activities, the repetition of commands every time we want to interact can be tedious, and result in the fragmentation of our daily tasks. 

\begin{figure}
\centering
  \includegraphics[width=0.9\columnwidth]{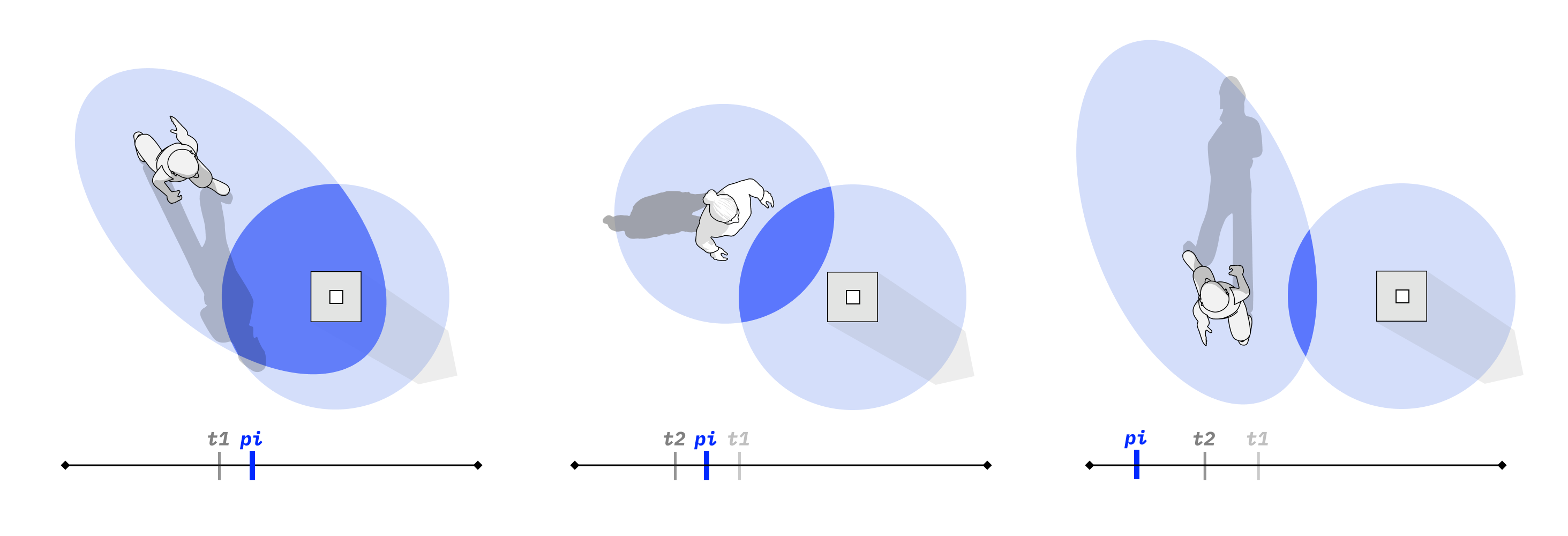}
  \caption{Greeting. Left: a user approaches a device and the potential interest value > t1. Middle: A user stand facing a device and potential interest value > t2. Right: A user walks away from a device and the potential interest value < t2.}
  \label{fig:greeting}
\end{figure}

The Greeting pattern defines an engagement threshold [t1], and changes its value based on the history of interaction, to seamlessly initiate, or end an interaction session. For example, when the Potential Interest value is above a defined threshold [t1], the device gets ready to interact with the user, switching from sleep to an active state: e.g. waking up the screen, turning on sensors such as touch-screen or microphone. Once the interaction with a device has been initiated, the engagement threshold can be lower [t2], making the dialogue with the device easier to maintain: e.g. even if the user moves a bit more outside the device interaction field, the device keeps an active state (Figure \ref{fig:greeting}) . Finally, when the Potential Interest value is lower than the engagement threshold [t2], the device goes back to a sleep state: it turns-off sensing systems which save power and preserve the user’s privacy.

This pattern can be used to accelerate frequent interactions with smart speakers and displays. In particular for voice-based interaction, it can also be used to disambiguate among a set of devices, which one the user is referring to when issuing a verbal command. It can be used also with laptops to increase security: for example, and as soon as the user leaves the area nearby the laptop, they are immediately logged off without the need of an explicit action. 

\subsection{Turn-taking}
The Turn-taking pattern allows a device to pause and resume an ongoing activity by understanding the amount of overlap between interaction fields, which is a byproduct of a user’s movements and body language. 

Particularly in our homes, the repetition of commands every time we want to initiate an interaction can be tedious, and result in the fragmentation of our daily tasks. This design pattern helps reduce repetitive tasks such as reaching out to play / pause videos and creates seamless transitions while a user is multitasking.

\begin{figure}
\centering
  \includegraphics[width=0.9\columnwidth]{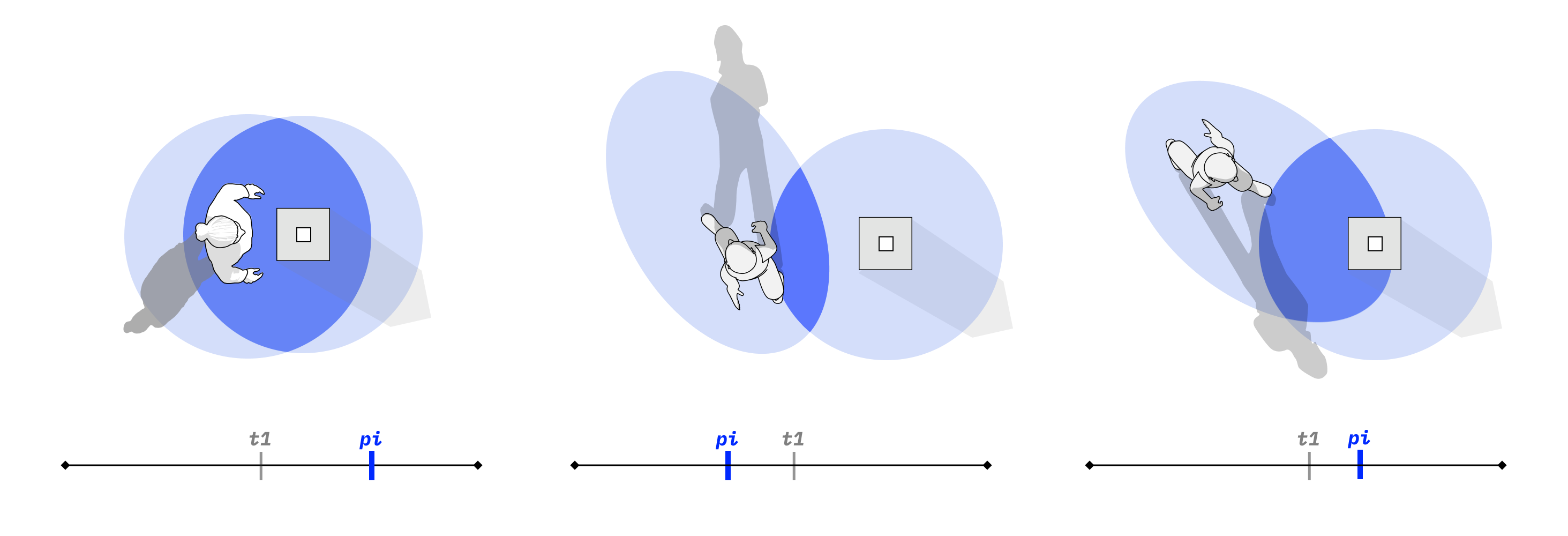}
  \caption Turn-taking. Left: a user is near device and the potential interest value > t1. Middle: A user turns away to grab an item and potential interest value < t1. Right: A user walks away from a device and the potential interest value > t1.
  \label{fig:turn}
\end{figure}

The Turn-taking pattern specifies a threshold [t1]. When an ongoing activity begins (such as playing a video) via touch, a user is engaged by default. Once the Potential Interest is lower than the engagement threshold [t1] the ongoing activity can pause. It only resumes once the Potential Interest is higher than the engagement threshold [t1] again (Figure \ref{fig:turn}). 

This pattern can be useful in a variety of tasks. One example is playing / pausing a recipe video. By understanding the potential interest to engage, a device can play and pause the video in the right moments. This pattern can also be used for more passive activities, like watching a video while someone is sitting on the couch. When a user gets up and leaves the device’s interaction field, the video can pause, and then resume when the Potential Interest is above a given threshold.

\subsection{Revealing}
\begin{figure}
\centering
  \includegraphics[width=0.9\columnwidth]{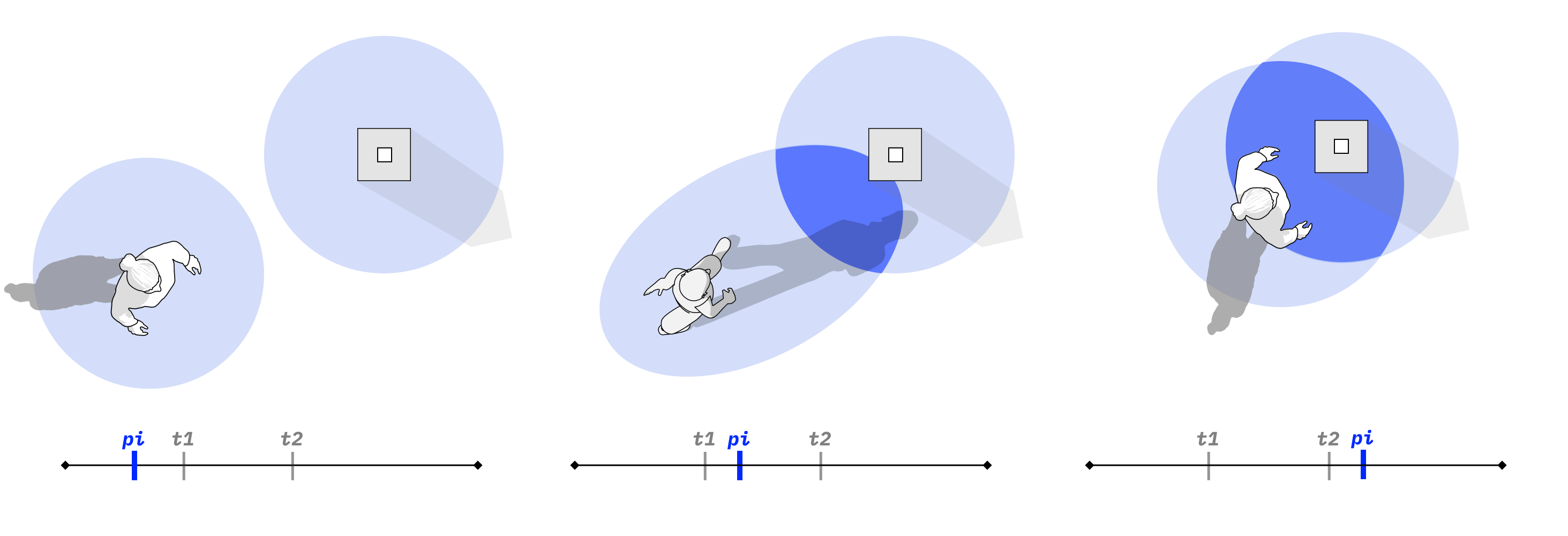}
  \caption{Revealing. Left: a user is near device and the potential interest is < t1. Middle: A user approaches and interest is > t1 but < 2. Right: A user stands near the device and the potential interest value > t2.}
  \label{fig:reveal}
\end{figure}

The Revealing allows a device to progressively reveal information based on a users’ awareness or engagement towards a device. 
Today, the majority of our interfaces are designed for touch interaction. As a result, many GUIs contain multiple pieces of information on a screen at once, and are primarily read from a close distance (within arm’s reach). This pattern provides more flexibility and customization for user interfaces to progressively reveal content based on different levels of awareness or engagement of a person relative to a device. 
The Progressive Reveal pattern typically specifies multiple thresholds [t1, t2, tx] that are corresponding to varying levels of information. When the Potential Interest is higher than the first threshold [t1], glanceable or minimal information can appear on an interface. When the Potential Interest is higher than the second threshold [t2], more detailed information can appear. This provides a situation in which two thresholds are used; theoretically, more thresholds can be added based on the context and information required (Figure \ref{fig:reveal}).
For example, a notification can show a preview while the device is in their periphery, showing just enough detail for them to notice it. When they are engaged with the device, a notification expands and shows more details and a possible call to action. 

This pattern can be used in a variety of contexts where a computing system needs to progressively disclose information. For example, this pattern can be used to create various levels of information density for a weather app. When a device is aware of a user’s presence, it can provide glanceable information such as the current temperature. As the user is engaged with the device, it can reveal a detailed view of the weekly weather forecast.

\section{Prototypes}
We built four prototypes to investigate potential applications of the Fields design framework.
These prototypes allowed us to explore the implementation of specific interaction patterns as defined in the previous section, and learn from user feedback about their desirability as well as emerging interaction design challenges.

\begin{figure}
\centering
  \includegraphics[width=0.9\columnwidth]{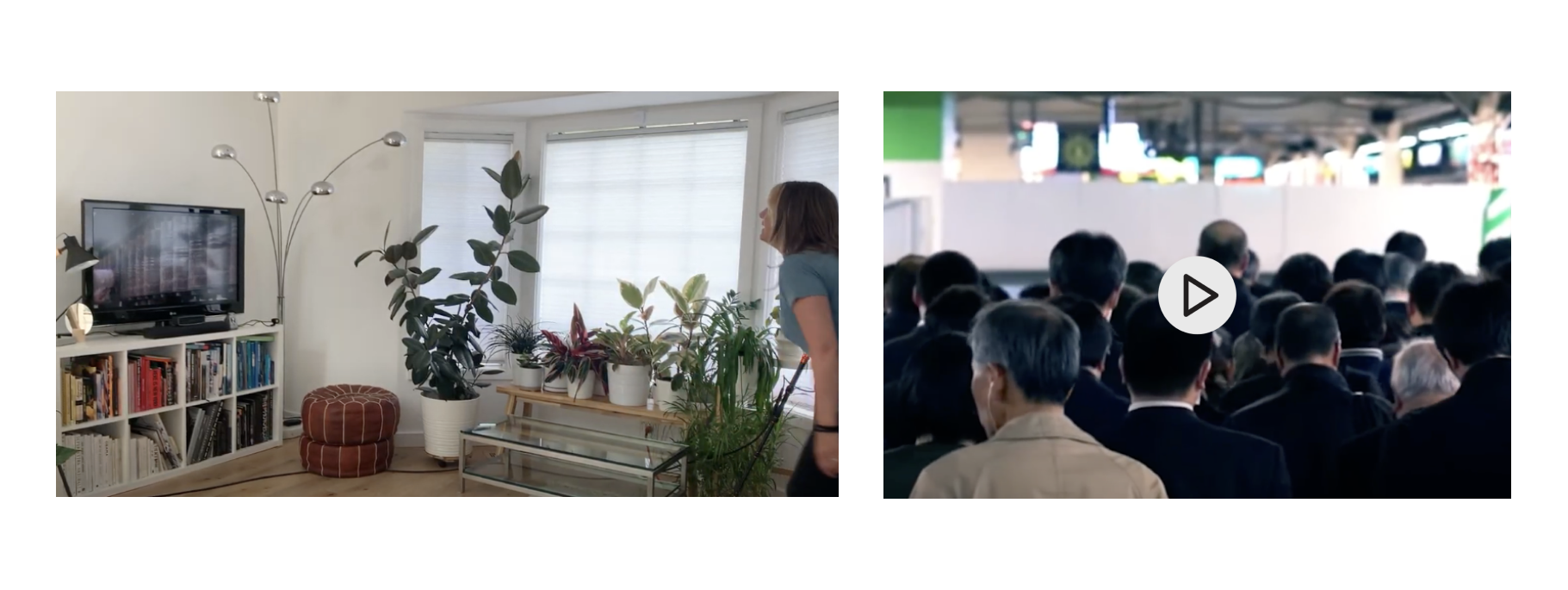}
  \caption{Play and Pause a Movie.  A video plays while participant is oriented in front of the screen, engaged with the video. The video pauses while participant leaves the area and is no longer engaged with the screen.
  \label{fig:caseone}
}
\end{figure}

\textbf{Play and Pause a Movie:} In this prototype, a video plays on a television: when the participant walks away for brief tasks, the video pauses. When the participant returns to the couch, the video resumes. This prototype implemented the turn-taking interaction pattern (Figure \ref{fig:caseone}).

\textbf{Email Notification:} On a tablet the system displays three levels of notification for a priority email. When the user is at a distance and orientation (say, walking away) connoting lack of engagement, no notification is displayed. At an intermediate level, a condensed notification with a subject appears. At a closer approach, the notification displays more detail, including a preview of the first part of the message. This use case investigates the use of the Revealing interaction pattern (Figure \ref{fig:casetwo}). 

\textbf{Scroll by Voice:} This use case prototype investigates how the Greeting interaction pattern can be used to follow a recipe step-by-step with a voice assistant, without repeating the wake word for every single interaction. When the user is engaged, the voice assistant wakes up (and the microphone is turned on) so that the user can ask to go to the next step (Figure \ref{fig:casethree}).

\begin{figure}
\centering
  \includegraphics[width=0.9\columnwidth]{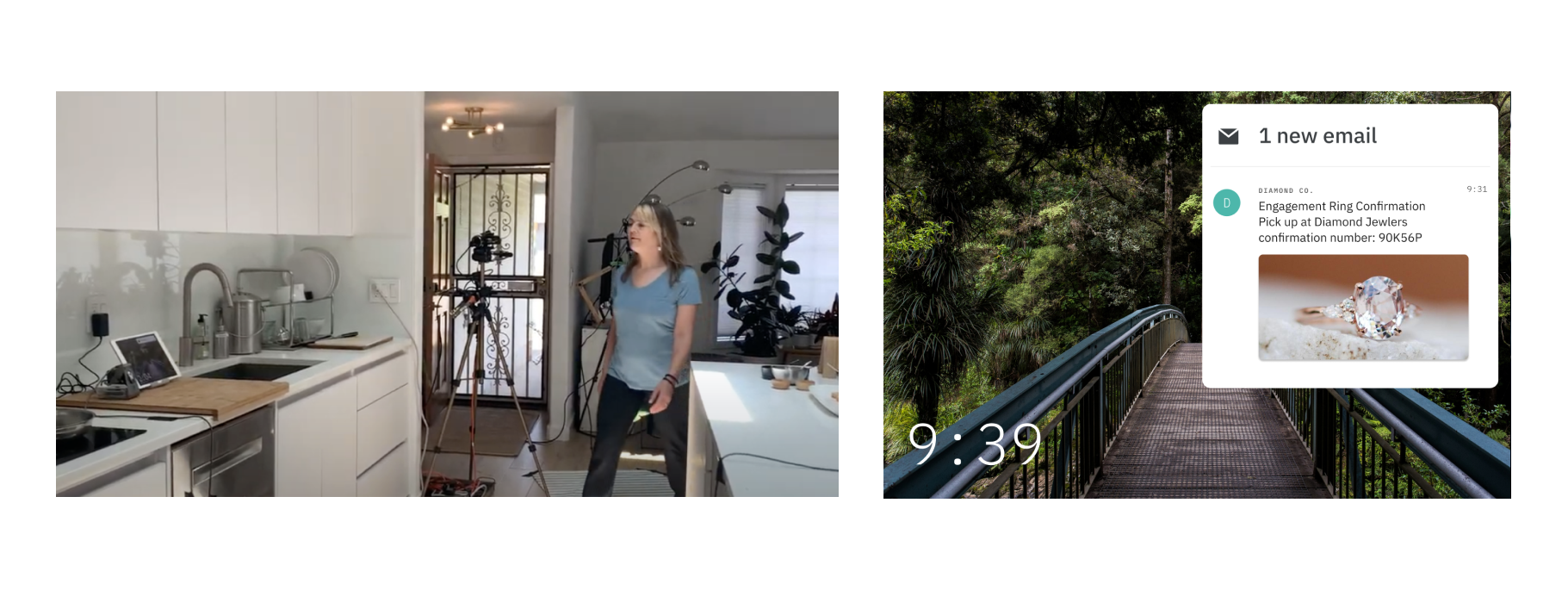}
  \caption{Email Notification. When the user is present, a small notification appears. When the user is engaged, the notification expands.
  \label{fig:casefour}}
\end{figure}

\begin{figure}
\centering
  \includegraphics[width=0.9\columnwidth]{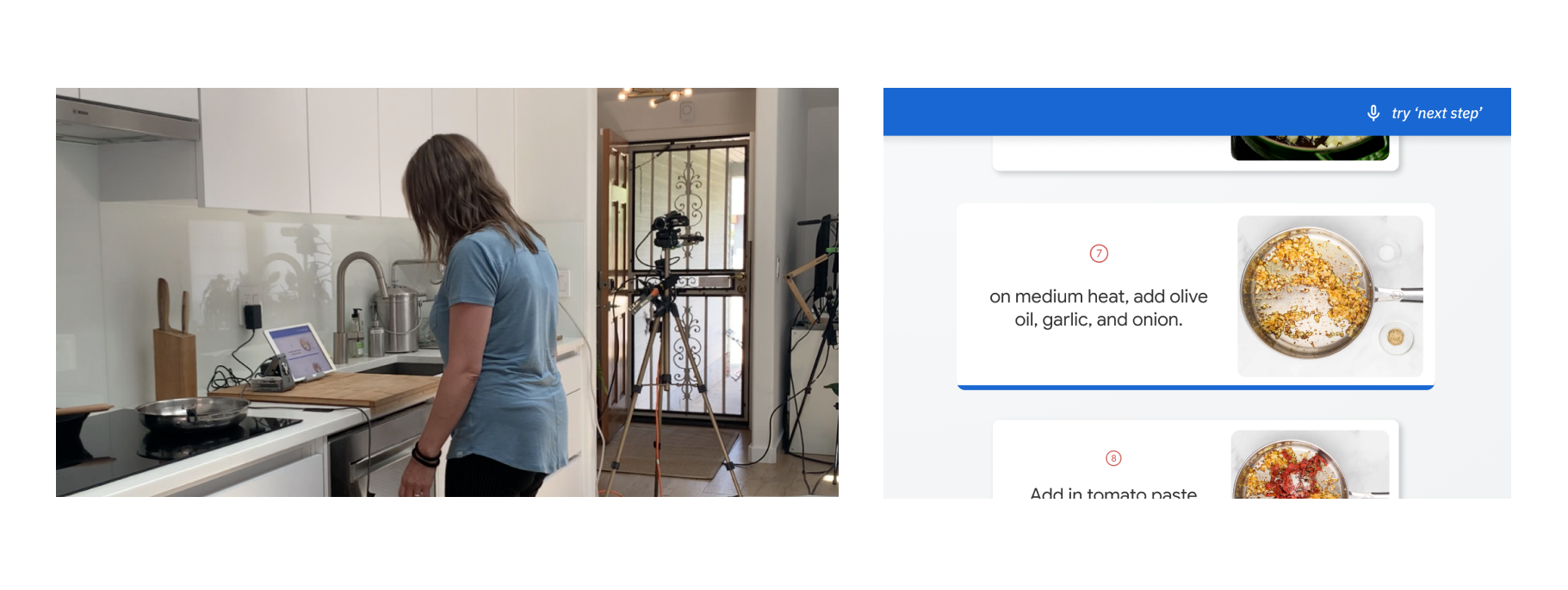}
  \caption{Scroll by Voice. When a user is engaged, the mic opens which option to advance to the next step of the recipe using a voice command. When the user is not engaged, the mic is not ‘open’, advancing to the next step requires a hotword.
  \label{fig:casethree}
}
\end{figure}

\begin{figure}
\centering
  \includegraphics[width=0.9\columnwidth]{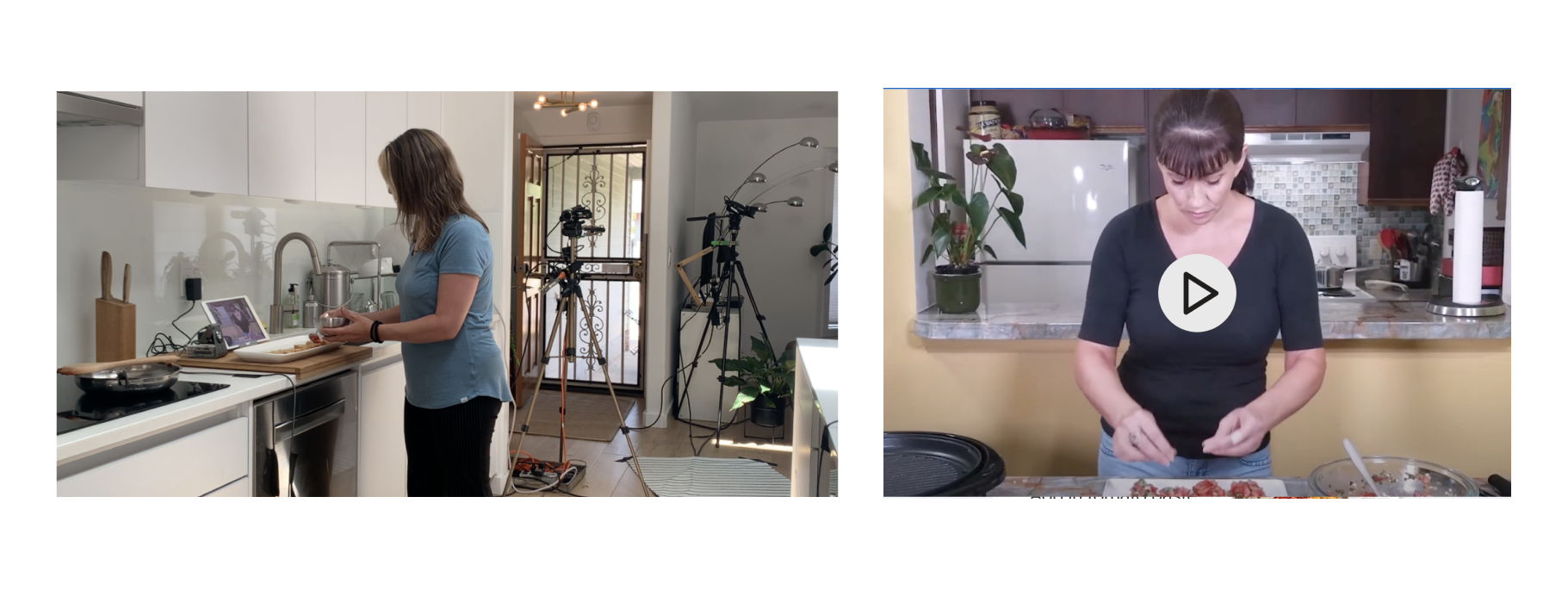}
  \caption{Play and Pause a Recipe A video plays while the user is engaged with the recipe. The video pauses while the user turns away from the device and is no longer engaged with the screen.
  \label{fig:casetwo}
  }
\end{figure}

\textbf{Play and Pause Recipe:} This prototypes also investigates the use of turn-taking interaction pattern, in the more dynamic context of the kitchen. A tablet playing a cooking how-to video pauses when the user moves away from the screen to grab a tool or an ingredient, then continues playback when the user moves back towards the device to continue cooking (Figure \ref{fig:casefour}).

\subsection{Implementation}
\begin{figure}
\centering
  \includegraphics[width=0.9\columnwidth]{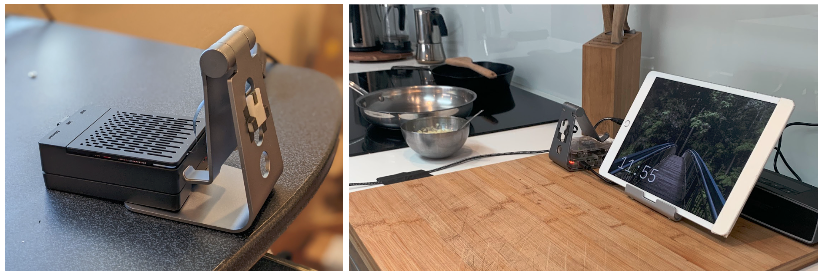}
  \caption{Left: We used a radar sensor connected to a Raspberry Pi 4 running a long-range presence algorithm that interprets sensor data and outputs inferred data about user movement. Right: We displayed the user interface for each use case on a tablet.
  \label{fig:proto}
}
\end{figure}

The Field framework we proposed in this paper can be supported by any technologies providing users’ and devices’ locations in physical spaces. We used Soli, a frequency modulated continuous wave (FMCW) radar, connected to a Raspberry Pi (Figure \ref{fig:proto}). In the Raspberry Pi, we executed DNN based algorithm to track the user's two dimensional position and velocity. The system allowed us to detect a user's location accurately in 5 meter x 5 meter space. The mean absolute error of its range and angle estimates was 0.05 (SD=0.14) meters and 5.5 (SD=7.4) degree. The output from the DNN algorithm was processed with a heuristic algorithm that calculated the potential interest (Eq \ref{eq:iou}).

As discussed in Section 3, we implemented a user's field as a ellipse changing its shape as a function the user's velocity as shown in (Eq \ref{eq:ellipse1}) and (Eq  \ref{eq:ellipse2}), where $R_1$ and $R_2$ denote the major and minor axes of the ellipse, $v$ is the speed, and $k$ is a dynamics coefficient. The major axis is in the direction of $\boldsymbol{v}$.

Following the existing literature \cite{hall1966hidden} we used the radius of $R_1 and R_2=1.2$ meters when $v=0$ (i.e., $c=1.2^2$). Because all devices used in the prototype were display devices, we defined the devices fields as a semicircle with a radius varying based on different prototypes (Table \ref{tab:quant}). In the first two prototypes, we empirically chose $k=0.25$, and for the latter two prototypes we used $k=0.5$ For the .25 k value, we found this responds better to courser changes of body position, while the .5 k value is more sensitive, and is more reactive to subtler changes of orientation, such as turning towards and away from the device. Based on the layout of the room and the task at hand, we chose which movements to optimize for (turning, or changes in 2D position). Finally we set a Potential Interest thresholds as shown in Table (Table \ref{tab:quant}).For the Revealing interaction pattern used in the Email notification we set the two as a basic starting point for a first evaluation. 

\begin{align}
    \frac{R_1}{R_2} &= kv + 1     \label{eq:ellipse1}\\
    R_1R_2 &= c
    \label{eq:ellipse2}
\end{align}

 \begin{table}
    \centering
    \caption{Summary of the values defined for each prototype. UIF = User Interaction Field; DIF = Device Interaction Field; Thresholds}
    \begin{tabular}{l r r r r}
        {\small\textit{Use Case}}
        & {\small \textit{UIF Radius}}
         & {\small \textit{UIF K Coefficient}}
         & {\small \textit{DIF Radius}}
         & {\small \textit{ Thresholds}}\\
        \midrule
        Play/pause entertainment video & 1.2m & .25 & 3m & .14 \\
        Email notification & 1.2m & .25 & 4m & .04, .08 \\
        Scroll by voice & 1.2m & .5 & 1m & .6\\
        Play/pause recipe video & 1.2m & .5 & 1m & .6 \\
        
    \end{tabular}
    \label{tab:quant}
\end{table}

\section{Evaluation}
We conducted a lab study of the four prototypes to evaluate their desirability, and study their impact on everyday tasks, such as cooking following recipes on smart displays or watching TV, in terms of percieved effort and ease of focus. We tested the prototypes with 14 participants  who were already daily users of smart home devices. We combined the advantages of lab study consistency with the advantages of a realistic, in-home usage environment by conducting all sessions in the kitchen and living room of a single fully-furnished, rented home.

\subsection{Participants}
Recruited participants were daily users of smart home devices such as a home assistant, smart thermostat, smart camera doorbell, smart door lock, Wi-Fi-enabled light bulb, or smart home security system, with a mix of gender identity, age, and ethnicity. Of the 14 participants who completed study sessions, 43 percent were female and 57 percent were male. Age ranges were 7 percent 18-24, 43 percent 25-34, 43 percent 35-44, and 7 perfect 45-54. Ethnicities were 21 percent Asian or Pacific islander, 7 percent Hispanic, 7 percent Middle Eastern, 7 percent Native American, 42 percent White, and 14 percent other.

\subsection{Protocol}
During recruiting, participants completed a questionnaire about demographics and smart home and mobile device usage. 

Test sessions were held in a rented home, where the living room served as a lobby and testing of use cases took place in the kitchen and living room. Each session began with the moderator giving a short description of the radar technology and what it would do. Then participants were invited to spend several minutes playing with a “use case 0” prototype that showed a real-time visualization of the level of engagement computed by an algorithm. We had some concern about participants having too much of an “oh, wow” experience and thus focusing more on the novelty of the technology than on the use case experiences. To mitigate that effect, we wanted to demystify the technology, so we created a use case 00 prototype for participants to try before experiencing any of the main use cases.

Next, participants experienced the four main use cases—play/pause entertainment video, email notification, scroll recipe instructions by voice, and play/pause recipe video—in one of 8 different sequences to reduce order effects. Each use case was designed to exercise the functionality several times, giving participants a solid opportunity to experience the technology. 

Following each use case, the moderator asked the participant to rate their experience on two 5-point scales:

\begin{itemize}
    \item \textbf{Effort:} This rating captures the level of effort required to carry out the task compared to a traditional device, where 1 means a standard smartphone or tablet requires less effort and 5 means the prototype they just used requires less effort.
    \item \textbf{Ease of focus:} This rating captures how easy it was to focus on their task or goal compared to traditional devices, essentially a distraction measurement, where 1 means a standard smartphone or tablet allows them to focus more on their task, and a 5 means the prototype allows them to focus more on their task.
\end{itemize}

The moderator also probed for participant rationales for their ratings and gathered free-form comments on the attraction of the use case.
The moderator then collected a rating on the value of this interaction capability overall, encouraging them to think of the capability generally, e.g., not limited to cooking or TV-watching.
A final set of open-ended questions was designed to give participants multiple ways of articulating their reaction to the technology. The moderator asked participants to talk about how the use case experiences compared to a traditional smart device, what the experience felt like, how they would describe the technology to a friend, what impact these kinds of interactions had on the way they think about technology, and so on.

\subsection{Quantitative Results}
All quantitative ratings are on a -2 to +2 scale, where a positive score is better for the prototype technology. Ratings from the 1 to 5 scale were converted to a zero-center scale by subtracting 3 from each score. 
Ratings for low effort (i.e., less need for tapping, swiping, speaking) compared to traditional devices and focus (i.e., freedom from distraction) compared to traditional devices for each use case are shown in the charts below, with a maximum mean of +2.0 for each and error bars showing the standard error.
A score of 0.0 would mean this technology was considered equally effortless and allowed easy focus on the task vs. a traditional smart device. A +2.0 score would mean this new technology was definitely more effortless and easier to focus with. -2.0 would mean a traditional smart device was definitely better. Mean scores for all four use cases were on the positive side for the prototype, even the least popular ones (Table \ref{tab:quant1}).

\begin{table}
    \centering
    \caption{Ratings for low effort (i.e., less need for tapping, swiping, speaking) compared to traditional devices and focus (i.e., freedom from distraction) compared to traditional devices for each use case are shown in the charts below, with a maximum mean of +2.0 for each and error bars showing the standard error.}
    \begin{tabular}{l r r}
        {\small\textit{Use Case}}
        & {\small \textit{Effortless}}
          & {\small \textit{Easy to Focus}}\\
        \midrule
        Play/pause entertainment video & +1.4 & +1.3 \\
        Email notification & +1.0 & +0.5 \\
        Scroll by voice & +1.2 & +1.5\\
        Play/pause recipe video & +0.9 & +0.5 \\
        Mean for all use cases & +1.1 & +1.0
    \end{tabular}
    \label{tab:quant1}
\end{table}

\begin{figure}
\centering
  \includegraphics[width=0.5\columnwidth]{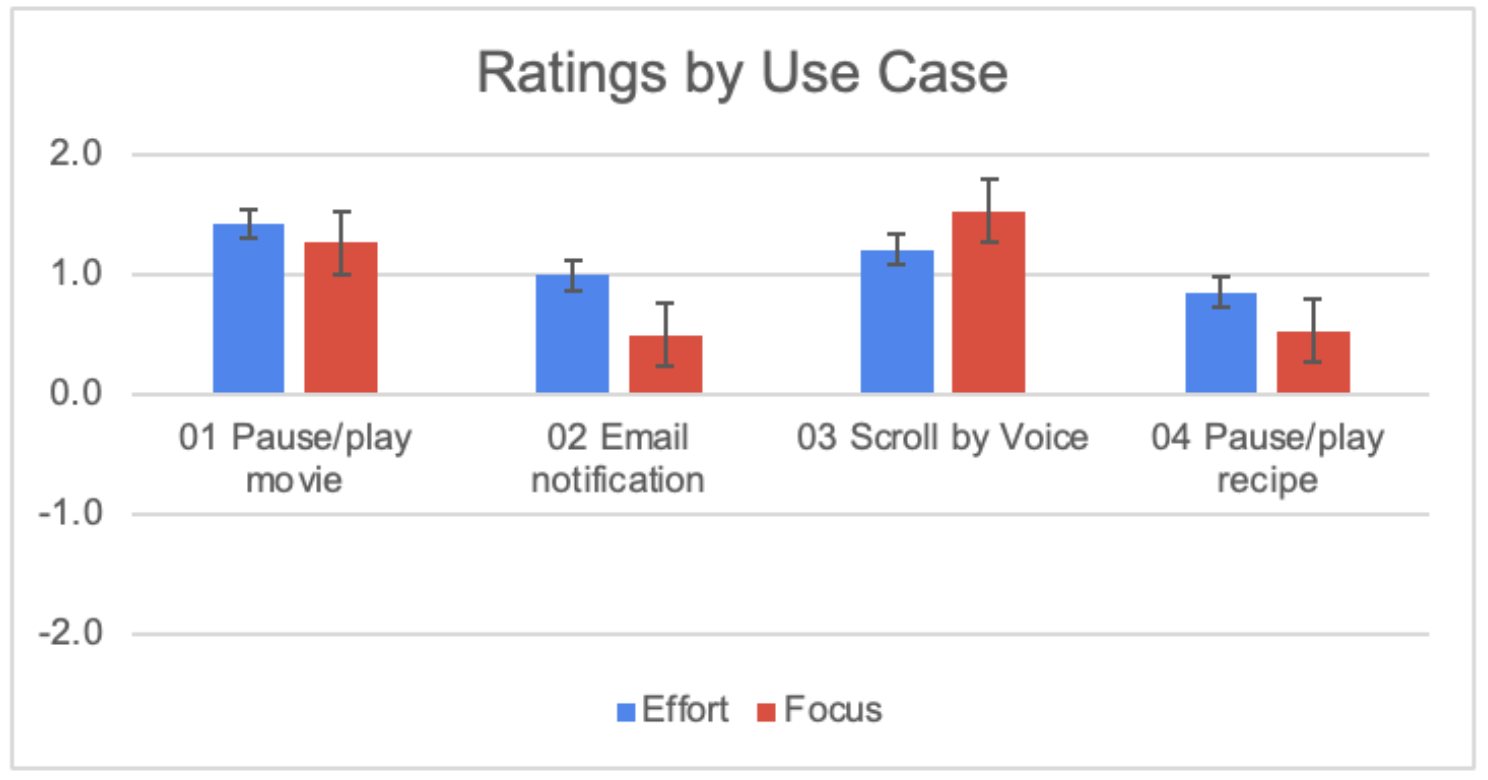}
  \caption  Ratings on Effort and Focus across all four use cases. Mean scores for all four use cases were on the positive side for the prototype, even the least popular ones.
  \label{fig:allcases}
\end{figure}

These ratings prefigure the top-2 rankings that show the play/pause entertainment video and scroll by voice use cases well ahead of the other two use cases. They also show that what is difficult about the less popular use cases is something related to focus, which the verbal explanations given by participants make clear.
These four use cases are quite different in some respects: for example, the recipe scroll prompt use case merely displays a message that makes a voice control invocation unnecessary, rather than directly taking an action on behalf of the user. Nonetheless, it may be worthwhile to average ratings across use cases to get a sense of the overall value of this technology, shown in figure 6. The mean rating overall was +1.1 for effort and +1.0 for focus, a positive result even when including the least attractive use cases (Figure \ref{fig:allcases}).

Finally, as a measure of overall value, participants rated how much they would want this general capability in the next smart home device they purchase on a 5-point Likert scale, where 1 is “I’m not interested in having that capability” and 5 is “I’m very interested in having that capability.” Figure 8 shows those results. No participants gave the technology a rating of less than 3, and 12 of 14 participants chose the highest ratings (4 or 5), for a mean rating of 4.3 out of 5.

\subsection{Qualitative Results}
In our analysis we looked especially closely at rationales given for effort and focus ratings, as well as freeform comments made by participants as they tried out the prototypes and later described their experiences in the final debrief discussion.

\begin{itemize}
    \item \textbf{Play/pause entertainment video:} In quantitative ratings, this use case received high scores for both effort and focus. Why? When asked for rationales for their scores, participants gave similar reasons. Typical answers about the positive value focused on obviating the need to pause using a remote (which requires three or four steps each time). Comments included these: “I didn’t have to do anything special.” “I think it’s fascinating, that idea of not having to search for a remote.” “If you’re watching something and you need to get out, you don’t need to look for the remote.” “The device I just used is definitely easier.” “I don’t have to do anything, right? I can still like grab whatever I want to grab and go back and [it plays] for me. I’m not distracted by the remote.” 
    \item \textbf{Email notification:} Participants explained that modest ratings for email notification were a result not of the technology but of the nature of notifications. First, people already get too many notifications of emails, messages, etc., so having another trigger was considered too much. Second, they worried about other household members walking by and seeing their message previews. One representative response summed it up: “It may be easier, but it’s not preferable.” Indeed, only 2 of the 14 participant scores for effort favored traditional smart devices, but for ability to focus on their task, 5 of 14 did. 
    \item \textbf{Scroll by voice:} The scroll by voice use case was second highest in the top-2 exercise and had a higher focus score than the most popular use case (play/pause entertainment video). Verbal comments showed that what led to high scores was the practical value. People said that being able to advance recipe instructions when their hands have food on them was highly convenient. Furthermore, they had full control over when to scroll the screen, rather than having it happen automatically when they approached the device, which might not match their need in the moment. Most people assumed the prompt was shown all the time, and very few noticed they had skipped the invocation. The moderator’s task introduction did explain how the software for this use case worked, but this information was quickly forgotten. 
    \item \textbf{Play/pause recipe video:} This use case has an interesting overlap with two others: it has identical functionality with the play/pause entertainment video use case, i.e., engagement triggers play or pause of a video; and it overlaps with the scroll by voice use case in that both allowed advancing through how-to content for cooking. The critical difference was not the device functionality; it was the task context. As one participant said, “My only concern with that was if I’m standing in front of it and it’s going, I’m starting to fall behind or something like that. I didn’t try [and see] if there’s a way to pause.” A secondary issue was that because cooking requires mental focus and frequent movement in a constrained space, sudden starting and stopping of the video became distracting. These issues led to to middling and lower ratings for focus.
\end{itemize}

Study participants summarized the potential of the technology in mainly positive descriptions. “The fact I didn’t have to prompt something to be a smart device was really convenient. I could just go about whatever I’m doing.” “It’s exciting. It’s effortless.” “The convenience factor feels like night and day to me [compared to] a voice-activated thing, which is most of my smart devices.” These comments align with the overall ratings, where participants’ desire to have the technology in their next smart home device was high, a mean of 4.3 out of 5.

The few negative terms tended toward hedging, with positives mixed in: “good but limited,” “choppy at times,” “nice but not necessary,” “takes getting used to.” One participant commented, “I definitely have a lot of concerns but that doesn’t mean that I wouldn't be totally gung ho about trying it.” Another said, “I think it could be very useful…unless it gets it wrong.”

\subsection{Discussion}
As shown by the variation in reactions to the four use case experiences, it is difficult to get user feedback on an interaction framework without having participants chiefly evaluate the particular implementation. We mitigated this effect by testing the prototypes in  a variety of contexts: we tested them in two of the rooms mostly likely to contain smart home devices, the kitchen and the living room, covering a variety of use cases: passive entertainment (play/pause entertainment video), instructional text (scroll by voice), how-to video (play/pause recipe video), and communications (email notification). In spite of the different responses to the prototypes, we can extrapolate some important learning regarding the Field design framework.

Overall, the user feedback across use cases came through positively. Ratings on the effort savings compared to traditional smart devices were averaging +1.1 (on a -2 to +2 scale) across all use cases; similarly the ease of focus scored +1.0 on average. This seems to suggest that proposed interaction design framework and its implementation was successfully in helping user accomplish a variety of tasks with less effort and more focus than traditional devices. It is interesting to notice how the common denominator of the least attractive use cases, Email Notification and Play and Pause Recipe was the misalignment with how people normally handle these tasks. While the Fields framework can help reduce effort and increase focus for certain tasks, it is important to situate its implementation in the specific context of use, for example taking into consideration privacy concerns for notifications, or the natural flow of everyday activities, such as cooking. 

It is interesting to observe how people describe the whole experience. For example, one participant said: “It’s only when I’m paying attention to it that it’s paying attention to me.” Another one, said “It’s nice that you don’t have to initiate it yourself, that it’s basically the one initiating it". While this requires more investigation, these comments seem to refer to a user's mental model of the system, where both the user and the device play an active role in establishing the conditions for the interaction to occur. And overall, participants felt that the prototypes were participating in the task execution in a way that felt more considerate. As one participant put it: “I think it requires a lot less intentional effort. [...] It’s nice to be able to just move around and it kind of responds to you like an ebb and flow, in a natural way.” In the same vein: "It adapts to what you are doing to make it easier for you to do certain things without having to be really connected to your devices". 

However, while people responded positively to the reduction of explicit input, they were concerned that the device wouldn't recognize their specific needs. A common theme for participants, mentioned across the different use cases, was finding themselves pulled between the urge to customize and the urge to step back: they wanted to make the technology do just what they wanted when they wanted, but they also wanted to relax and let it help them. One put it this way: “Obviously I would want it to work exactly how I want it to, but I think in the end what I would prefer is having something that’s more automated and I don’t have to think about what I’m doing.” Another looked ahead to having more smart home technology: “I think maybe in the future it would be more necessary when we do have more technology around us—you don’t want to have to fiddle with every single little [device]. So if it is able to do things on its own for us, that would be nice.” In the implementation of the Fields framework is therefore important to consider the level of customization left to the end-users. Design and developers can make specific choices in relation to the shape of the interaction field, and specific Potential Interest thresholds, given a certain application and context of use; however, there might be certain situation where the user might want to control certain aspects of the implementation to fine-tune the overall system behaviors to their unique needs.  

In conclusion, both numerical subjective ratings and verbal comments by participants suggest that the Fields framework is a potential approach for the design of ubiquitous systems that reduce the need of user's input and attention, while supporting the execution of everyday tasks. More in general, the qualitative results seem to suggest that this approach can lead to the implementation of devices that are able to participate in our everyday life in a way that feels more considerate and natural.

\section{Limitations and Future work}
In this paper, we introduced Fields, a framework to inform the design of connected products with social grace. Inspired by interactionist theories, the Fields framework builds on the idea that the physical space we share with computers can be an interface to mediate interactions. It defines a generalized approach to design for spatial interactions along with a set of interaction patterns that can be adapted to a variety of computing systems. We investigated its value by implementing it in a set of prototypes, which were evaluated with end users in a lab setting, simulating a living spaces (e.g. kitchen, living room). 

The results of our evaluation demonstrate the use case prototypes came through positively. This suggests the proposed interaction design framework and its implementation was successful in helping user accomplish a variety of tasks with less effort and more focus than traditional devices. Additionally, our qualitative analysis and general remarks on the technology provides early hints at a shift of mental model that points to devices being helpful and considerate. As one participant put it: “I think it requires a lot less intentional effort. [...] It’s nice to be able to just move around and it kind of responds to you like an ebb and flow, in a natural way.”  The remarks behind these prototypes also suggest receptiveness to a new paradigm where devices are equal participants in the interaction cycle. 

There are a few limitations worth noting for this work. First, we defined the Fields framework with a basic representation of space, and more sophisticated parameters can be provided to allow designer and developers to explore different shapes and configurations. Second, the algorithm used in this study is still very early, and not ready yet to be deployed in a real world scenario. This naturally affected the evaluation, constraining it to a lab setting. Furthermore, since the prototypes and interaction concepts presented were novel to users, and participants only had a limited amount of time to try out the use cases, it is reasonable to think that ratings and comments would change if the participant had more time to experience the technology. 

These limitations suggest a few areas for future work. The first is to deepen our work on the framework itself. One area includes adding additional flexibility in how we consider interaction fields. For example, In the IF framework we can add flexibility to adopt the notion of “field” to be more than a traditional in-out zone. Interaction fields can have an homogenous magnitude, with a clear definition of their boundaries; or they can be defined as gradient or density function. 

Another area of future work expand this framework to accommodate several interaction typologies, including how multiple people and multiple devices. For example, what is the significance of multiple fields overlapping? What if multiple people's interaction fields overlap? What if someone is wearing a device? This opens up new areas of spatial and social context understanding. For example, if the interaction fields of two people overlap while in front of a TV, perhaps suggestions appear that are relevant for group settings. Additionally, if two people are watching a movie and one person leaves, perhaps the TV bookmarks the place where a user left off. In addition, we'd like to build additional prototypes for different use cases that test a range of different parameters (such as changes in the size of the device interaction field) and prototypes that do not contain screens.  Since ubiquitous computing is grounded in multi-user and multi-device systems, these contexts seems to be a valuable next step.

We'd also like to develop an expanded set of prototypes that help refine our interaction pattern library, our use of thresholds, and even serve as a vehicle to develop new interaction patterns. With these prototypes, we would like to add additional rounds of evaluation on a wider variety of use cases that, taken together, give participants a broader sense of potential applicability, as befits a test seeking to evaluate a technology. We'd also like to have more robust prototypes that can support longitudinal studies. We do see benefit in participants having more exploration time and a longer experience with each use case to see the nuances of interaction. 

While this is just a first step, we believe that an approach like Fields has the potential to become a new standard within interaction design for ubiquitous computing systems. In the same way interface guidelines specify appropriate ways to use inputs (mouse and keyboard) and interaction patterns to standardize user experience (cursors, hover states, window managers), we believe that a new set of interaction patterns and standards will have to emerge as we move computing "off-desktop" and into the physical world. Fields, we believe, pushes the design practice towards this direction as a starting point. We hope this work facilitates continued discussion toward spatial interaction standards for ubiquitous computing. 


\bibliographystyle{ACM-Reference-Format}
\bibliography{field}


\begin{thebibliography}{46}


\ifx \showCODEN    \undefined \def \showCODEN     #1{\unskip}     \fi
\ifx \showDOI      \undefined \def \showDOI       #1{#1}\fi
\ifx \showISBNx    \undefined \def \showISBNx     #1{\unskip}     \fi
\ifx \showISBNxiii \undefined \def \showISBNxiii  #1{\unskip}     \fi
\ifx \showISSN     \undefined \def \showISSN      #1{\unskip}     \fi
\ifx \showLCCN     \undefined \def \showLCCN      #1{\unskip}     \fi
\ifx \shownote     \undefined \def \shownote      #1{#1}          \fi
\ifx \showarticletitle \undefined \def \showarticletitle #1{#1}   \fi
\ifx \showURL      \undefined \def \showURL       {\relax}        \fi
\providecommand\bibfield[2]{#2}
\providecommand\bibinfo[2]{#2}
\providecommand\natexlab[1]{#1}
\providecommand\showeprint[2][]{arXiv:#2}

\bibitem[\protect\citeauthoryear{Abowd and Mynatt}{Abowd and Mynatt}{2000}]%
        {abowd2000charting}
\bibfield{author}{\bibinfo{person}{Gregory~D Abowd} {and}
  \bibinfo{person}{Elizabeth~D Mynatt}.} \bibinfo{year}{2000}\natexlab{}.
\newblock \showarticletitle{Charting past, present, and future research in
  ubiquitous computing}.
\newblock \bibinfo{journal}{\emph{ACM Transactions on Computer-Human
  Interaction (TOCHI)}} \bibinfo{volume}{7}, \bibinfo{number}{1}
  (\bibinfo{year}{2000}), \bibinfo{pages}{29--58}.
\newblock


\bibitem[\protect\citeauthoryear{Abowd, Mynatt, and Rodden}{Abowd
  et~al\mbox{.}}{2002}]%
        {abowd2002human}
\bibfield{author}{\bibinfo{person}{Gregory~D Abowd},
  \bibinfo{person}{Elizabeth~D Mynatt}, {and} \bibinfo{person}{Tom Rodden}.}
  \bibinfo{year}{2002}\natexlab{}.
\newblock \showarticletitle{The human experience [of ubiquitous computing]}.
\newblock \bibinfo{journal}{\emph{IEEE pervasive computing}}
  \bibinfo{volume}{1}, \bibinfo{number}{1} (\bibinfo{year}{2002}),
  \bibinfo{pages}{48--57}.
\newblock


\bibitem[\protect\citeauthoryear{Ahire and Rohs}{Ahire and Rohs}{2020}]%
        {ahire2020tired}
\bibfield{author}{\bibinfo{person}{Shashank Ahire} {and}
  \bibinfo{person}{Michael Rohs}.} \bibinfo{year}{2020}\natexlab{}.
\newblock \showarticletitle{Tired of wake words? Moving towards seamless
  conversations with intelligent personal assistants}. In
  \bibinfo{booktitle}{\emph{Proceedings of the 2nd Conference on Conversational
  User Interfaces}}. \bibinfo{pages}{1--3}.
\newblock


\bibitem[\protect\citeauthoryear{Anderson, H{\"u}bener, Seipp, Ohly, David, and
  Pejovic}{Anderson et~al\mbox{.}}{2018}]%
        {anderson2018survey}
\bibfield{author}{\bibinfo{person}{Christoph Anderson}, \bibinfo{person}{Isabel
  H{\"u}bener}, \bibinfo{person}{Ann-Kathrin Seipp}, \bibinfo{person}{Sandra
  Ohly}, \bibinfo{person}{Klaus David}, {and} \bibinfo{person}{Veljko
  Pejovic}.} \bibinfo{year}{2018}\natexlab{}.
\newblock \showarticletitle{A survey of attention management systems in
  ubiquitous computing environments}.
\newblock \bibinfo{journal}{\emph{Proceedings of the ACM on Interactive,
  Mobile, Wearable and Ubiquitous Technologies}} \bibinfo{volume}{2},
  \bibinfo{number}{2} (\bibinfo{year}{2018}), \bibinfo{pages}{1--27}.
\newblock


\bibitem[\protect\citeauthoryear{Bailey and Konstan}{Bailey and
  Konstan}{2006}]%
        {bailey2006need}
\bibfield{author}{\bibinfo{person}{Brian~P Bailey} {and}
  \bibinfo{person}{Joseph~A Konstan}.} \bibinfo{year}{2006}\natexlab{}.
\newblock \showarticletitle{On the need for attention-aware systems: Measuring
  effects of interruption on task performance, error rate, and affective
  state}.
\newblock \bibinfo{journal}{\emph{Computers in human behavior}}
  \bibinfo{volume}{22}, \bibinfo{number}{4} (\bibinfo{year}{2006}),
  \bibinfo{pages}{685--708}.
\newblock


\bibitem[\protect\citeauthoryear{Ballendat, Marquardt, and Greenberg}{Ballendat
  et~al\mbox{.}}{2010}]%
        {ballendat2010proxemic}
\bibfield{author}{\bibinfo{person}{Till Ballendat}, \bibinfo{person}{Nicolai
  Marquardt}, {and} \bibinfo{person}{Saul Greenberg}.}
  \bibinfo{year}{2010}\natexlab{}.
\newblock \showarticletitle{Proxemic interaction: designing for a proximity and
  orientation-aware environment}. In \bibinfo{booktitle}{\emph{ACM
  International Conference on Interactive Tabletops and Surfaces}}.
  \bibinfo{pages}{121--130}.
\newblock


\bibitem[\protect\citeauthoryear{Blumer}{Blumer}{1986}]%
        {blumer1986symbolic}
\bibfield{author}{\bibinfo{person}{Herbert Blumer}.}
  \bibinfo{year}{1986}\natexlab{}.
\newblock \bibinfo{booktitle}{\emph{Symbolic interactionism: Perspective and
  method}}.
\newblock \bibinfo{publisher}{Univ of California Press}.
\newblock


\bibitem[\protect\citeauthoryear{Bufacchi and Iannetti}{Bufacchi and
  Iannetti}{2018}]%
        {bufacchi2018action}
\bibfield{author}{\bibinfo{person}{Rory~J Bufacchi} {and}
  \bibinfo{person}{Gian~Domenico Iannetti}.} \bibinfo{year}{2018}\natexlab{}.
\newblock \showarticletitle{An action field theory of peripersonal space}.
\newblock \bibinfo{journal}{\emph{Trends in cognitive sciences}}
  \bibinfo{volume}{22}, \bibinfo{number}{12} (\bibinfo{year}{2018}),
  \bibinfo{pages}{1076--1090}.
\newblock


\bibitem[\protect\citeauthoryear{Clark}{Clark}{1996}]%
        {clark1996arranging}
\bibfield{author}{\bibinfo{person}{Herbert~H Clark}.}
  \bibinfo{year}{1996}\natexlab{}.
\newblock \showarticletitle{Arranging to do things with others}. In
  \bibinfo{booktitle}{\emph{Conference Companion on Human Factors in Computing
  Systems}}. \bibinfo{pages}{165--167}.
\newblock


\bibitem[\protect\citeauthoryear{Clark, Pantidi, Cooney, Doyle, Garaialde,
  Edwards, Spillane, Gilmartin, Murad, Munteanu, et~al\mbox{.}}{Clark
  et~al\mbox{.}}{2019}]%
        {clark2019makes}
\bibfield{author}{\bibinfo{person}{Leigh Clark}, \bibinfo{person}{Nadia
  Pantidi}, \bibinfo{person}{Orla Cooney}, \bibinfo{person}{Philip Doyle},
  \bibinfo{person}{Diego Garaialde}, \bibinfo{person}{Justin Edwards},
  \bibinfo{person}{Brendan Spillane}, \bibinfo{person}{Emer Gilmartin},
  \bibinfo{person}{Christine Murad}, \bibinfo{person}{Cosmin Munteanu},
  {et~al\mbox{.}}} \bibinfo{year}{2019}\natexlab{}.
\newblock \showarticletitle{What makes a good conversation? Challenges in
  designing truly conversational agents}. In
  \bibinfo{booktitle}{\emph{Proceedings of the 2019 CHI Conference on Human
  Factors in Computing Systems}}. \bibinfo{pages}{1--12}.
\newblock


\bibitem[\protect\citeauthoryear{Coello and Cartaud}{Coello and
  Cartaud}{2021}]%
        {coello2021interrelation}
\bibfield{author}{\bibinfo{person}{Yann Coello} {and} \bibinfo{person}{Alice
  Cartaud}.} \bibinfo{year}{2021}\natexlab{}.
\newblock \showarticletitle{The interrelation between peripersonal action space
  and interpersonal social space: psychophysiological evidence and clinical
  implications}.
\newblock \bibinfo{journal}{\emph{Frontiers in Human Neuroscience}}
  \bibinfo{volume}{15} (\bibinfo{year}{2021}), \bibinfo{pages}{636124}.
\newblock


\bibitem[\protect\citeauthoryear{Dey}{Dey}{2001}]%
        {dey2001understanding}
\bibfield{author}{\bibinfo{person}{Anind~K Dey}.}
  \bibinfo{year}{2001}\natexlab{}.
\newblock \showarticletitle{Understanding and using context}.
\newblock \bibinfo{journal}{\emph{Personal and ubiquitous computing}}
  \bibinfo{volume}{5}, \bibinfo{number}{1} (\bibinfo{year}{2001}),
  \bibinfo{pages}{4--7}.
\newblock


\bibitem[\protect\citeauthoryear{Fruin}{Fruin}{1992}]%
        {fruin1992designing}
\bibfield{author}{\bibinfo{person}{John Fruin}.}
  \bibinfo{year}{1992}\natexlab{}.
\newblock \showarticletitle{Designing for pedestrians}.
\newblock \bibinfo{journal}{\emph{Public Transportation United States}}
  (\bibinfo{year}{1992}).
\newblock


\bibitem[\protect\citeauthoryear{Fujinami}{Fujinami}{2010}]%
        {fujinami2010interaction}
\bibfield{author}{\bibinfo{person}{Kaori Fujinami}.}
  \bibinfo{year}{2010}\natexlab{}.
\newblock \showarticletitle{Interaction design issues in smart home
  environments}. In \bibinfo{booktitle}{\emph{2010 5th International Conference
  on Future Information Technology}}. IEEE, \bibinfo{pages}{1--8}.
\newblock


\bibitem[\protect\citeauthoryear{Gershenfeld}{Gershenfeld}{1999}]%
        {gershenfeld1999things}
\bibfield{author}{\bibinfo{person}{Neil Gershenfeld}.}
  \bibinfo{year}{1999}\natexlab{}.
\newblock \bibinfo{booktitle}{\emph{When things start to think}}.
\newblock \bibinfo{publisher}{Macmillan}.
\newblock


\bibitem[\protect\citeauthoryear{Gibbs}{Gibbs}{2005}]%
        {gibbs2005considerate}
\bibfield{author}{\bibinfo{person}{W~Wayt Gibbs}.}
  \bibinfo{year}{2005}\natexlab{}.
\newblock \showarticletitle{Considerate computing}.
\newblock \bibinfo{journal}{\emph{Scientific American}} \bibinfo{volume}{292},
  \bibinfo{number}{1} (\bibinfo{year}{2005}), \bibinfo{pages}{54--61}.
\newblock


\bibitem[\protect\citeauthoryear{Goffman}{Goffman}{1967}]%
        {goffman1967interaction}
\bibfield{author}{\bibinfo{person}{Erving Goffman}.}
  \bibinfo{year}{1967}\natexlab{}.
\newblock \showarticletitle{Interaction ritual: Essays on face-to-face
  interaction.}
\newblock  (\bibinfo{year}{1967}).
\newblock


\bibitem[\protect\citeauthoryear{Greenberg, Marquardt, Ballendat, Diaz-Marino,
  and Wang}{Greenberg et~al\mbox{.}}{2011}]%
        {greenberg2011proxemic}
\bibfield{author}{\bibinfo{person}{Saul Greenberg}, \bibinfo{person}{Nicolai
  Marquardt}, \bibinfo{person}{Till Ballendat}, \bibinfo{person}{Rob
  Diaz-Marino}, {and} \bibinfo{person}{Miaosen Wang}.}
  \bibinfo{year}{2011}\natexlab{}.
\newblock \showarticletitle{Proxemic interactions: the new ubicomp?}
\newblock \bibinfo{journal}{\emph{interactions}} \bibinfo{volume}{18},
  \bibinfo{number}{1} (\bibinfo{year}{2011}), \bibinfo{pages}{42--50}.
\newblock


\bibitem[\protect\citeauthoryear{Hall and Hall}{Hall and Hall}{1966}]%
        {hall1966hidden}
\bibfield{author}{\bibinfo{person}{Edmund~T Hall} {and}
  \bibinfo{person}{Edward~Twitchell Hall}.} \bibinfo{year}{1966}\natexlab{}.
\newblock \bibinfo{booktitle}{\emph{The hidden dimension}}.
  Vol.~\bibinfo{volume}{609}.
\newblock \bibinfo{publisher}{Anchor}.
\newblock


\bibitem[\protect\citeauthoryear{Hong and Suh}{Hong and Suh}{[n.d.]}]%
        {hong2009kim}
\bibfield{author}{\bibinfo{person}{Jong-yi Hong} {and} \bibinfo{person}{Eui-ho
  Suh}.} \bibinfo{year}{[n.d.]}\natexlab{}.
\newblock \showarticletitle{Kim. S.(2009). Context-aware systems}.
\newblock \bibinfo{journal}{\emph{Expert Syst. Appl}} \bibinfo{volume}{36},
  \bibinfo{number}{4} (\bibinfo{year}{[n.\,d.]}), \bibinfo{pages}{8509--8522}.
\newblock


\bibitem[\protect\citeauthoryear{Ju, Lee, and Klemmer}{Ju
  et~al\mbox{.}}{2008}]%
        {ju2008range}
\bibfield{author}{\bibinfo{person}{Wendy Ju}, \bibinfo{person}{Brian~A Lee},
  {and} \bibinfo{person}{Scott~R Klemmer}.} \bibinfo{year}{2008}\natexlab{}.
\newblock \showarticletitle{Range: exploring implicit interaction through
  electronic whiteboard design}. In \bibinfo{booktitle}{\emph{Proceedings of
  the 2008 ACM conference on Computer supported cooperative work}}.
  \bibinfo{pages}{17--26}.
\newblock


\bibitem[\protect\citeauthoryear{Ju and Leifer}{Ju and Leifer}{2008}]%
        {ju2008design}
\bibfield{author}{\bibinfo{person}{Wendy Ju} {and} \bibinfo{person}{Larry
  Leifer}.} \bibinfo{year}{2008}\natexlab{}.
\newblock \showarticletitle{The design of implicit interactions: Making
  interactive systems less obnoxious}.
\newblock \bibinfo{journal}{\emph{Design Issues}} \bibinfo{volume}{24},
  \bibinfo{number}{3} (\bibinfo{year}{2008}), \bibinfo{pages}{72--84}.
\newblock


\bibitem[\protect\citeauthoryear{Kendon}{Kendon}{1990}]%
        {kendon1990conducting}
\bibfield{author}{\bibinfo{person}{Adam Kendon}.}
  \bibinfo{year}{1990}\natexlab{}.
\newblock \bibinfo{booktitle}{\emph{Conducting interaction: Patterns of
  behavior in focused encounters}}. Vol.~\bibinfo{volume}{7}.
\newblock \bibinfo{publisher}{CUP Archive}.
\newblock


\bibitem[\protect\citeauthoryear{Koetsier}{Koetsier}{2022}]%
        {koetsier_2022}
\bibfield{author}{\bibinfo{person}{John Koetsier}.}
  \bibinfo{year}{2022}\natexlab{}.
\newblock \bibinfo{title}{Smart home: Apple is the fastest-growing connected
  device company}.
\newblock
\newblock
\urldef\tempurl%
\url{https://www.forbes.com/sites/johnkoetsier/2022/08/31/smart-home-apple-is-the-fastest-growing-connected-device-company/?sh=38920817dd48}
\showURL{%
\tempurl}


\bibitem[\protect\citeauthoryear{Leroy}{Leroy}{2009}]%
        {leroy2009so}
\bibfield{author}{\bibinfo{person}{Sophie Leroy}.}
  \bibinfo{year}{2009}\natexlab{}.
\newblock \showarticletitle{Why is it so hard to do my work? The challenge of
  attention residue when switching between work tasks}.
\newblock \bibinfo{journal}{\emph{Organizational Behavior and Human Decision
  Processes}} \bibinfo{volume}{109}, \bibinfo{number}{2}
  (\bibinfo{year}{2009}), \bibinfo{pages}{168--181}.
\newblock


\bibitem[\protect\citeauthoryear{Lloyd}{Lloyd}{2009}]%
        {lloyd2009space}
\bibfield{author}{\bibinfo{person}{Donna~M Lloyd}.}
  \bibinfo{year}{2009}\natexlab{}.
\newblock \showarticletitle{The space between us: A neurophilosophical
  framework for the investigation of human interpersonal space}.
\newblock \bibinfo{journal}{\emph{Neuroscience \& Biobehavioral Reviews}}
  \bibinfo{volume}{33}, \bibinfo{number}{3} (\bibinfo{year}{2009}),
  \bibinfo{pages}{297--304}.
\newblock


\bibitem[\protect\citeauthoryear{Lyytinen and Yoo}{Lyytinen and Yoo}{2002}]%
        {lyytinen2002ubiquitous}
\bibfield{author}{\bibinfo{person}{Kalle Lyytinen} {and}
  \bibinfo{person}{Youngjin Yoo}.} \bibinfo{year}{2002}\natexlab{}.
\newblock \showarticletitle{Ubiquitous computing}.
\newblock \bibinfo{journal}{\emph{Commun. ACM}} \bibinfo{volume}{45},
  \bibinfo{number}{12} (\bibinfo{year}{2002}), \bibinfo{pages}{63--96}.
\newblock


\bibitem[\protect\citeauthoryear{Lyytinen, Yoo, Varshney, Ackerman, Davis,
  Avital, Robey, Sawyer, and Sorensen}{Lyytinen et~al\mbox{.}}{2004}]%
        {lyytinen2004surfing}
\bibfield{author}{\bibinfo{person}{Kalle~J Lyytinen}, \bibinfo{person}{Youngjin
  Yoo}, \bibinfo{person}{Upkar Varshney}, \bibinfo{person}{Mark Ackerman},
  \bibinfo{person}{Gordon Davis}, \bibinfo{person}{Michel Avital},
  \bibinfo{person}{Daniel Robey}, \bibinfo{person}{Steve Sawyer}, {and}
  \bibinfo{person}{Carsten Sorensen}.} \bibinfo{year}{2004}\natexlab{}.
\newblock \showarticletitle{Surfing the next wave: design and implementation
  challenges of ubiquitous computing}.
\newblock \bibinfo{journal}{\emph{Communications of the Association for
  information systems}} \bibinfo{volume}{13}, \bibinfo{number}{1}
  (\bibinfo{year}{2004}), \bibinfo{pages}{40}.
\newblock


\bibitem[\protect\citeauthoryear{Mark, Gudith, and Klocke}{Mark
  et~al\mbox{.}}{2008}]%
        {mark2008cost}
\bibfield{author}{\bibinfo{person}{Gloria Mark}, \bibinfo{person}{Daniela
  Gudith}, {and} \bibinfo{person}{Ulrich Klocke}.}
  \bibinfo{year}{2008}\natexlab{}.
\newblock \showarticletitle{The cost of interrupted work: more speed and
  stress}. In \bibinfo{booktitle}{\emph{Proceedings of the SIGCHI conference on
  Human Factors in Computing Systems}}. \bibinfo{pages}{107--110}.
\newblock


\bibitem[\protect\citeauthoryear{Marquardt, Diaz-Marino, Boring, and
  Greenberg}{Marquardt et~al\mbox{.}}{2011}]%
        {marquardt2011proximity}
\bibfield{author}{\bibinfo{person}{Nicolai Marquardt}, \bibinfo{person}{Robert
  Diaz-Marino}, \bibinfo{person}{Sebastian Boring}, {and} \bibinfo{person}{Saul
  Greenberg}.} \bibinfo{year}{2011}\natexlab{}.
\newblock \showarticletitle{The proximity toolkit: prototyping proxemic
  interactions in ubiquitous computing ecologies}. In
  \bibinfo{booktitle}{\emph{Proceedings of the 24th annual ACM symposium on
  User interface software and technology}}. \bibinfo{pages}{315--326}.
\newblock


\bibitem[\protect\citeauthoryear{Marquardt and Greenberg}{Marquardt and
  Greenberg}{2015}]%
        {marquardt2015proxemic}
\bibfield{author}{\bibinfo{person}{Nicolai Marquardt} {and}
  \bibinfo{person}{Saul Greenberg}.} \bibinfo{year}{2015}\natexlab{}.
\newblock \showarticletitle{Proxemic interactions: From theory to practice}.
\newblock \bibinfo{journal}{\emph{Synthesis Lectures on Human-Centered
  Informatics}} \bibinfo{volume}{8}, \bibinfo{number}{1}
  (\bibinfo{year}{2015}), \bibinfo{pages}{1--199}.
\newblock


\bibitem[\protect\citeauthoryear{Marquardt, Hinckley, and Greenberg}{Marquardt
  et~al\mbox{.}}{2012}]%
        {marquardt2012cross}
\bibfield{author}{\bibinfo{person}{Nicolai Marquardt}, \bibinfo{person}{Ken
  Hinckley}, {and} \bibinfo{person}{Saul Greenberg}.}
  \bibinfo{year}{2012}\natexlab{}.
\newblock \showarticletitle{Cross-device interaction via micro-mobility and
  f-formations}. In \bibinfo{booktitle}{\emph{Proceedings of the 25th annual
  ACM symposium on User interface software and technology}}.
  \bibinfo{pages}{13--22}.
\newblock


\bibitem[\protect\citeauthoryear{Muller, Alt, Michelis, and Schmidt}{Muller
  et~al\mbox{.}}{2010}]%
        {muller2010requirements}
\bibfield{author}{\bibinfo{person}{Jorg Muller}, \bibinfo{person}{Florian Alt},
  \bibinfo{person}{Daniel Michelis}, {and} \bibinfo{person}{Albrecht Schmidt}.}
  \bibinfo{year}{2010}\natexlab{}.
\newblock \showarticletitle{Requirements and design space for interactive
  public displays}. In \bibinfo{booktitle}{\emph{Proceedings of the 18th ACM
  international conference on Multimedia}}. \bibinfo{pages}{1285--1294}.
\newblock


\bibitem[\protect\citeauthoryear{Persson}{Persson}{2018}]%
        {persson2018framing}
\bibfield{author}{\bibinfo{person}{Anders Persson}.}
  \bibinfo{year}{2018}\natexlab{}.
\newblock \bibinfo{booktitle}{\emph{Framing social interaction: Continuities
  and cracks in Goffman’s Frame Analysis}}.
\newblock \bibinfo{publisher}{Taylor and Francis}.
\newblock


\bibitem[\protect\citeauthoryear{Prante, R{\"o}cker, Streitz, Stenzel,
  Magerkurth, Van~Alphen, and Plewe}{Prante et~al\mbox{.}}{2003}]%
        {prante2003hello}
\bibfield{author}{\bibinfo{person}{Thorsten Prante}, \bibinfo{person}{Carsten
  R{\"o}cker}, \bibinfo{person}{Norbert Streitz}, \bibinfo{person}{Richard
  Stenzel}, \bibinfo{person}{Carsten Magerkurth}, \bibinfo{person}{Daniel
  Van~Alphen}, {and} \bibinfo{person}{Daniela Plewe}.}
  \bibinfo{year}{2003}\natexlab{}.
\newblock \showarticletitle{Hello. wall--beyond ambient displays}. In
  \bibinfo{booktitle}{\emph{Adjunct Proceedings of Ubicomp}},
  Vol.~\bibinfo{volume}{2003}. \bibinfo{pages}{277--278}.
\newblock


\bibitem[\protect\citeauthoryear{R{\"a}dle, Jetter, Marquardt, Reiterer, and
  Rogers}{R{\"a}dle et~al\mbox{.}}{2014}]%
        {radle2014huddlelamp}
\bibfield{author}{\bibinfo{person}{Roman R{\"a}dle},
  \bibinfo{person}{Hans-Christian Jetter}, \bibinfo{person}{Nicolai Marquardt},
  \bibinfo{person}{Harald Reiterer}, {and} \bibinfo{person}{Yvonne Rogers}.}
  \bibinfo{year}{2014}\natexlab{}.
\newblock \showarticletitle{Huddlelamp: Spatially-aware mobile displays for
  ad-hoc around-the-table collaboration}. In
  \bibinfo{booktitle}{\emph{Proceedings of the Ninth ACM International
  Conference on Interactive Tabletops and Surfaces}}. \bibinfo{pages}{45--54}.
\newblock


\bibitem[\protect\citeauthoryear{Rizzolatti, Fadiga, Fogassi, and
  Gallese}{Rizzolatti et~al\mbox{.}}{1997}]%
        {rizzolatti1997space}
\bibfield{author}{\bibinfo{person}{Giacomo Rizzolatti},
  \bibinfo{person}{Luciano Fadiga}, \bibinfo{person}{Leonardo Fogassi}, {and}
  \bibinfo{person}{Vittorio Gallese}.} \bibinfo{year}{1997}\natexlab{}.
\newblock \showarticletitle{The space around us}.
\newblock \bibinfo{journal}{\emph{Science}} \bibinfo{volume}{277},
  \bibinfo{number}{5323} (\bibinfo{year}{1997}), \bibinfo{pages}{190--191}.
\newblock


\bibitem[\protect\citeauthoryear{Roussel, Evans, and Hansen}{Roussel
  et~al\mbox{.}}{2004}]%
        {roussel2004proximity}
\bibfield{author}{\bibinfo{person}{Nicolas Roussel}, \bibinfo{person}{Helen
  Evans}, {and} \bibinfo{person}{Heiko Hansen}.}
  \bibinfo{year}{2004}\natexlab{}.
\newblock \showarticletitle{Proximity as an interface for video communication}.
\newblock \bibinfo{journal}{\emph{IEEE MultiMedia}} \bibinfo{volume}{11},
  \bibinfo{number}{3} (\bibinfo{year}{2004}), \bibinfo{pages}{12--16}.
\newblock


\bibitem[\protect\citeauthoryear{Stryker and Vryan}{Stryker and Vryan}{2006}]%
        {stryker2006symbolic}
\bibfield{author}{\bibinfo{person}{Sheldon Stryker} {and}
  \bibinfo{person}{Kevin~D Vryan}.} \bibinfo{year}{2006}\natexlab{}.
\newblock \showarticletitle{The symbolic interactionist frame}.
\newblock \bibinfo{journal}{\emph{Handbook of social psychology}}
  (\bibinfo{year}{2006}), \bibinfo{pages}{3--28}.
\newblock


\bibitem[\protect\citeauthoryear{Teneggi, Canzoneri, di~Pellegrino, and
  Serino}{Teneggi et~al\mbox{.}}{2013}]%
        {teneggi2013social}
\bibfield{author}{\bibinfo{person}{Chiara Teneggi}, \bibinfo{person}{Elisa
  Canzoneri}, \bibinfo{person}{Giuseppe di Pellegrino}, {and}
  \bibinfo{person}{Andrea Serino}.} \bibinfo{year}{2013}\natexlab{}.
\newblock \showarticletitle{Social modulation of peripersonal space
  boundaries}.
\newblock \bibinfo{journal}{\emph{Current biology}} \bibinfo{volume}{23},
  \bibinfo{number}{5} (\bibinfo{year}{2013}), \bibinfo{pages}{406--411}.
\newblock


\bibitem[\protect\citeauthoryear{Vertegaal et~al\mbox{.}}{Vertegaal
  et~al\mbox{.}}{2003}]%
        {vertegaal2003attentive}
\bibfield{author}{\bibinfo{person}{Roel Vertegaal} {et~al\mbox{.}}}
  \bibinfo{year}{2003}\natexlab{}.
\newblock \showarticletitle{Attentive user interfaces}.
\newblock \bibinfo{journal}{\emph{Commun. ACM}} \bibinfo{volume}{46},
  \bibinfo{number}{3} (\bibinfo{year}{2003}), \bibinfo{pages}{30--33}.
\newblock


\bibitem[\protect\citeauthoryear{Vogel and Balakrishnan}{Vogel and
  Balakrishnan}{2004}]%
        {vogel2004interactive}
\bibfield{author}{\bibinfo{person}{Daniel Vogel} {and} \bibinfo{person}{Ravin
  Balakrishnan}.} \bibinfo{year}{2004}\natexlab{}.
\newblock \showarticletitle{Interactive public ambient displays: transitioning
  from implicit to explicit, public to personal, interaction with multiple
  users}. In \bibinfo{booktitle}{\emph{Proceedings of the 17th annual ACM
  symposium on User interface software and technology}}.
  \bibinfo{pages}{137--146}.
\newblock


\bibitem[\protect\citeauthoryear{Wang, Boring, and Greenberg}{Wang
  et~al\mbox{.}}{2012}]%
        {wang2012proxemic}
\bibfield{author}{\bibinfo{person}{Miaosen Wang}, \bibinfo{person}{Sebastian
  Boring}, {and} \bibinfo{person}{Saul Greenberg}.}
  \bibinfo{year}{2012}\natexlab{}.
\newblock \showarticletitle{Proxemic peddler: a public advertising display that
  captures and preserves the attention of a passerby}. In
  \bibinfo{booktitle}{\emph{Proceedings of the 2012 international symposium on
  pervasive displays}}. \bibinfo{pages}{1--6}.
\newblock


\bibitem[\protect\citeauthoryear{Weiser}{Weiser}{1991}]%
        {weiser1991computer}
\bibfield{author}{\bibinfo{person}{Mark Weiser}.}
  \bibinfo{year}{1991}\natexlab{}.
\newblock \showarticletitle{The Computer for the 21st Century}.
\newblock \bibinfo{journal}{\emph{Scientific american}} \bibinfo{volume}{265},
  \bibinfo{number}{3} (\bibinfo{year}{1991}), \bibinfo{pages}{94--105}.
\newblock


\bibitem[\protect\citeauthoryear{Weiser and Brown}{Weiser and Brown}{1996}]%
        {weiser1996designing}
\bibfield{author}{\bibinfo{person}{Mark Weiser} {and}
  \bibinfo{person}{John~Seely Brown}.} \bibinfo{year}{1996}\natexlab{}.
\newblock \showarticletitle{Designing calm technology}.
\newblock \bibinfo{journal}{\emph{PowerGrid Journal}} \bibinfo{volume}{1},
  \bibinfo{number}{1} (\bibinfo{year}{1996}), \bibinfo{pages}{75--85}.
\newblock


\bibitem[\protect\citeauthoryear{Weiser and Brown}{Weiser and Brown}{1997}]%
        {weiser1997coming}
\bibfield{author}{\bibinfo{person}{Mark Weiser} {and}
  \bibinfo{person}{John~Seely Brown}.} \bibinfo{year}{1997}\natexlab{}.
\newblock \showarticletitle{The coming age of calm technology}.
\newblock In \bibinfo{booktitle}{\emph{Beyond calculation}}.
  \bibinfo{publisher}{Springer}, \bibinfo{pages}{75--85}.
\newblock


\end{thebibliography}

\end{document}